\documentclass{article}

\usepackage{arxiv}
\usepackage{graphicx}
\usepackage{hhline}
\usepackage{pgf}
\usepackage{multirow}
\usepackage{subcaption}
\usepackage{siunitx}
\usepackage[utf8]{inputenc} % allow utf-8 input
\usepackage[T1]{fontenc}    % use 8-bit T1 fonts
\usepackage{hyperref}       % hyperlinks
\usepackage{url}            % simple URL typesetting
\usepackage{booktabs}       % professional-quality tables
\usepackage{amsfonts}       % blackboard math symbols
\usepackage{nicefrac}       % compact symbols for 1/2, etc.
\usepackage{microtype}      % microtypography
\usepackage{lipsum}
\usepackage[]{authblk}
\usepackage[format=hang]{caption}
\usepackage{placeins}

\usepackage[printonlyused, nohyperlinks, smaller]{acronym}
\usepackage{amsmath}

\title{Biological sex classification with structural MRI data shows increased misclassification in \acl{tw}}

\author[$\;\;$ 1,4]{Claas Flint \thanks{indicates that the authors contributed equally to the work and should be regarded as first authors.}}
\author[$\;*$ 1]{Katharina Förster}
\author[1]{Sophie A. Koser}
\author[5]{Carsten Konrad}
\author[2]{Pienie Zwitserlood}
\author[3]{Klaus Berger}
\author[3]{Marco~Hermesdorf}
\author[8]{Tilo Kircher}
\author[8]{Igor Nenadic}
\author[8]{Axel Krug}
\author[1,6,7]{Bernhard T. Baune}
\author[1]{Katharina Dohm}
\author[1]{Ronny Redlich}
\author[1]{Nils Opel}
\author[1]{Volker Arolt}
\author[1]{Tim Hahn}
\author[4]{Xiaoyi Jiang}
\author[$\;\;\ddagger$1]{Udo Dannlowski \thanks{indicates that the authors contributed equally to the work and should be regarded as senior authors.}}
\author[$\;\dagger$1]{Dominik Grotegerd}

\affil[1]{Department of Psychiatry and Psychotherapy, University of Münster, Albert Schweitzer-Campus 1, A9, 48149 Münster, Germany}
\affil[2]{Department of Psychology, University of Münster, Fliednerstraße 21, 48149 Münster, Germany}
\affil[3]{Department of Epidemiology and Social Medicine, University of Münster, Albert Schweitzer-Campus 1, G D3, 48149 Münster}
\affil[4]{Department of Computer Science, University of Münster, Einsteinstraße 62, 48149 Münster, Germany}
\affil[5]{Department of Psychiatry and Psychotherapy, Agaplesion Diakonieklinikum, 27356 Rotenburg, Germany}
\affil[6]{Department of Psychiatry, Melbourne Medical School, The University of Melbourne, Parkville, Australia}
\affil[7]{The Florey Institute of Neuroscience and Mental Health, The University of Melbourne, Parkville, Australia}
\affil[8]{Department of Psychiatry and Psychotherapy, University of Marburg, Germany}
\affil[$\ddagger$]{Corresponding author: Udo Dannlowski, Phone: \href{tel:+492518356610}{+49-251-83-56610}, Fax: \href{fax:+492518356612}{+49-251-8356612}, Email: \href{mailto:dannlow@uni-muenster.de}{dannlow@uni-muenster.de}}

\begin{document}
    \maketitle

    \begin{abstract}
        \acp*{Ti}\acused{ti} show brain structural alterations that differ from their biological sex as well as their perceived gender. To substantiate evidence that the brain structure of \acp{ti} differs from male and female, we use a combined multivariate and univariate approach. Gray matter segments resulting from voxel-based morphometry preprocessing of $N = 1753$ \ac{cg} healthy participants were used to train ($N=1402$) and validate (\SI{20}{\percent} hold-out; $N = 351$) a support-vector machine classifying the biological sex. As a second validation, we classified $N = 1104$ patients with depression. A third validation was performed using the matched \ac{cg} sample of the \acp{tw} application-sample. Subsequently, the classifier was applied to $N = 26$ \acp{tw}. Finally, we compared brain volumes of \ac{cg}-men, women and \acp{tw}-pre/post treatment (\acl{cht}) in a univariate analysis controlling for sexual orientation, age and total brain volume. The application of our biological sex classifier to the transgender sample resulted in a significantly lower \ac{tpr} (\ac{tpr}-male = \SI{56.0}{\percent}). The \ac{tpr} did not differ between \ac{cg}-individuals with (\ac{tpr}-male = \SI{86.9}{\percent}) and without depression (\ac{tpr}-male = \SI{88.5}{\percent}). The univariate analysis of the transgender application-sample revealed that \acp{tw}-pre/post treatment show brain structural differences from \ac{cg}-women and \ac{cg}-men in the putamen and insula, as well as the whole-brain analysis. Our results support the hypothesis that brain structure in \acp{tw} differs from brain structure of their biological sex (male) as well as their perceived gender (female). This finding substantiates evidence that \acp{ti} show specific brain structural alterations leading to a different pattern of brain structure than \ac{cg}-individuals. \acresetall
    \end{abstract}

% keywords can be removed

    \section{Introduction}
    Being transgender describes the stable feeling of belonging to the opposite rather than the biological sex assigned at birth, while the term \ac{cg} describes the feeling of coherence between biological sex and perceived gender.

    Although there is an ongoing social and political debate regarding the terms and phrases used to describe gender, little is known about how a divergence between biological sex and perceived gender emerges. A popular view is that sexual brain differentiation and body development diverge in \acp{ti}~\cite{Zhou1995}. Evidence for this comes from studies in female infants with congenital adrenal hyperplasia, who develop male playing behavior~\cite{Meyer-Bahlburg1996, Mathews2009}. Due to prenatally circulating testosterone, the brain of such female infants is structurally organized as a male brain, while their body development is female~\cite{Zhou1995, Meyer-Bahlburg1996, Mathews2009, Bao2011, VanGoozen1995}.

    Previous research provides extensive information on how brain structure differs as a function of biological sex. Briefly, local sex differences show higher gray matter volume in \ac{cg}-men, while the volume of limbic structures is particularly increased in \ac{cg}-women~\cite{Ruigrok2014}. However, sexual differentiation seems less prominent in the brain compared to physical appearance~\cite{Cahill2006, McCarthy2011, Spizzirri2018}. Hence, brains cannot easily be classified into dimorphic gender categories~\cite{Joel2015}.

    Few ROI-based approaches have studied how brain structure of \acp{ti} differs from \ac{cg}. Compared to \ac{cg}-men, \acl{tw} (biological sex male, perceived gender female, \ac{tw}) show structural alterations of areas associated with body perception. Brain structures that repeatedly showed alterations across multiple studies are the putamen~\cite{Luders2009} and the insula~\cite{Savic2011}. However, the alterations are highly heterogeneous in their direction and the reported studies only investigated individuals before \ac{cht}. Comparisons between \ac{tw}-pre/post-\ac{cht} with \ac{cg}-individuals again exhibited heterogeneous results~\cite{Spizzirri2018, Mueller2017, Altinay2019, Seiger2016, Nguyen2018, Nguyen2018a, WhiteHughto2016}. \ac{cht} in \ac{tw} combines treatment with anti-androgens and estradiol and is associated with region-specific structural alterations of the brain~\cite{Kranz2017} such as local volume and cortical thickness decreases~\cite{Seiger2016, Mueller2017a}. However, longitudinal studies are scarce and a recent large study did not find any differences between \ac{tw}-pre and post-\ac{cht}~\cite{Spizzirri2018, Nguyen2018}.

    Next to univariate analyses, multivariate analyses offer new insights into the similarities and differences between \ac{cg} and \acp{ti}~\cite{Hoekzema2015, Baldinger-Melich2019}. In contrast to univariate analysis, multivariate analysis does not focus on identifying mean differences between individuals rather than recognizing the discriminative patterns within the data applicable on an individual level. This may be utilized to subdivide data into broader categories, but also to identify cases that exhibit unusual patterns and cannot be categorized easily. This approach is particularly interesting for \acp{ti}, since they perceive a disparity between their gender and their biological sex. Hence, one could assume that they represent cases that exhibit unusual data patterns, e.g. hormone levels, personality traits or brain function and structure. Recent studies also show a variety of brain-structural differences between \acp{ti} and \ac{cg}-individuals. Thus, a univariate approach might not be suitable to clarify how \acp{ti} and \ac{cg}-individuals differ from each other structurally.
    Another methodological motivation for choosing multivariate techniques is that samples of transgender individuals are usually small. Using a multivariate approach trained and validated on larges samples of \ac{cg}-individuals and applied to \acp{ti} allows more valid conclusions about brain structural differences between \acp{ti} and \ac{cg}-individuals.

    Multivariate analyses have already been used to investigate whether \acp{ti} can be separated from \ac{cg}-individuals by their brain volumetric patterns~\cite{Hoekzema2015,Baldinger-Melich2019}. Both studies show decreased accuracy in biological sex classification in \acp{ti} compared to \ac{cg}-individuals. However, it has been recently criticized that classifiers trained with small sample sizes often lead to high accuracies, but low external validity~\cite{Varoquaux2018}. Hence, in contrast to previous studies, we trained and validated a biological sex classifier with large samples of \ac{cg}-participants without any psychiatric comorbidities. We then applied the classifier to a smaller sample of \ac{tw}. To ensure that observed misclassification is not caused or biased by psychiatric comorbidity, we performed a second validation of the classifier in an additional large validation-sample with MDD-patients. A third validation was performed in a matched \ac{cg} sample of the \ac{tw} application-sample, whose data were recorded at the same time and in the same scanner.
    Thus, an extensively greater generalizability is expected and therefore real-life applicability is enhanced.
    Our hypotheses for the multivariate analysis are:
    \begin{itemize}
        \item[(1)] The classifier trained on healthy \ac{cg}-participants shows significantly worse performance when applied to a sample of \ac{tw}.
        \item[(2)] The classifier trained on healthy \ac{cg}-participants performs equally well in a validation-sample of \ac{cg}-patients suffering from major depression.
    \end{itemize}
    Following our multivariate approach, we investigated local structural brain alterations in the putamen and the insula~\cite{Kranz2015, Burke2018, Spizzirri2018, Savic2011, Luders2009, Zubiaurre-Elorza2014}. Since \ac{tw} differ in brain structure from both \ac{cg}-men and –women, with \ac{tw} exhibiting lower volume in the putamen~\cite{Savic2011} and insula~\cite{Spizzirri2018} than \ac{cg}-men, but lower volume than \ac{cg}-women~\cite{Spizzirri2018, Simon2013, Luders2012}, we hypothesize that
    \begin{itemize}
        \item[(3)]    \ac{cg}-women show lower volume in comparison to \ac{cg}-men~\cite{Ruigrok2014}.
        \item[(4)]     \ac{tw}-pre and post-\ac{cht} show increased volume in comparison to \ac{cg}-women.
        \item[(5)]    \ac{tw}-pre and post-\ac{cht} show lower volume in comparison to \ac{cg}-men.
    \end{itemize}
    Since we expect \ac{cht} to be associated with a further feminization of brain structure and hence reduced volume, we hypothesize that
    \begin{itemize}
        \item[(6)]    \ac{tw}-pre-\ac{cht} show higher volume in comparison to \ac{tw}-post-\ac{cht}.
    \end{itemize}
    Being transgender describes the stable feeling of belonging to the opposite rather than the biological sex assigned at birth, while the term \ac{cg} describes the feeling of coherence between biological sex and perceived gender.

    \section{Materials and Method}
    To obtain and validate a predictor for biological sex based on structural MRI brain scans, we used three different samples, which purposes are briefly described here prior to sample characteristics: A classifier was trained on a large sample of \ac{cg}-individuals without any psychiatric disorder using a cross-validation procedure. An independent subsample randomly drawn in advance, served as the first validation set, to avoid overfitting (supplementary figure~\ref{fig:flow_chart_classification}). To rule out that depressive symptoms influence the performance of the predictor in our \ac{tw}-group, we used a second validation-sample with \ac{mdd}-patients. Next, the classifier was applied to data from \ac{tw}-individuals, and to a third validation group whose data were acquired at the same time and with the same scanner as the \ac{tw}-sample.

    \subsection{Data}

    \subsubsection{\acl*{Cg}\acused{cg} training sample and first validation set}
    The data from a sample of $N = 1753$ \ac{cg}-participants without any evidence of previous psychiatric disorders served as the basis for the training. History of psychiatric disorders was ruled out using the Structured Clinical Interview following DSM-IV criteria~\cite{Wittchen1997}. The participants were taken from three different cohorts: the \acl{mnc} (\acs{mnc}\acused{mnc}, $N = 666$~\cite{Dannlowski2015}), the \acl{bd} (\acs{bd}\acused{mnc}, $N = 434$~\cite{Teuber2017}) and the FOR2107 study ($N = 653$~\cite{Kircher2018, Vogelbacher2018}). Exclusion criteria for the \ac{mnc} were presence or history of major internal or neurological disorder, dependence on or recent abuse of alcohol or drugs, hypertension, and general MRI contraindications. \ac{bd} and FOR2107 have similar exclusion criteria; details are described in supplementary table~\ref{tab:descrip_stats_train_validation} and elsewhere~\cite{Teismann2014, Kircher2018}.

    \subsubsection{Second, clinical validation sample – patients suffering from \acf{mdd}}
    To exclude that potential differences in classification \ac{tpr} rate are due to comorbid depressive symptoms in \ac{tw}, data from a clinical sample ($N=1404$) of patients diagnosed with \ac{mdd} were used as second validation-sample. 450 \ac{mdd} patients exhibited psychiatric comorbidities such as anxiety disorders or substance abuse. Diagnoses were again verified with the Structural Clinical Interview according to DSM-IV criteria~\cite{Wittchen1997}. The \ac{mdd} sample consisted of $N=285$ participants from the \ac{mnc}, $N=591$ from the \ac{bd} study and $N=528$ from the FOR2107 study (supplementary table~\ref{tab:descrip_stats_train_validation}). Additional exclusion criteria were presence of bipolar disorder, schizoaffective disorders and schizophrenia, substance-related disorders, current benzodiazepine treatment (wash out of at least three half-lives before study participation), and recent electroconvulsive therapy. Nearly all patients were under psychopharmacological antidepressant treatment and/or received psychotherapy.

    \subsubsection{Application: transgender application sample including third validation sample}
    To test for a different classification of \ac{cg} and \ac{tw}, we used an independent sample of $N=29$ \ac{tw}. Three \ac{tw} had to be excluded from our analysis due to poor image quality and artifacts. Data of \ac{tw} were collected in conjunction with a set of \ac{cg}-controls that serve as the third validation-sample of $N=19$ \ac{cg}-women and $N=15$ \ac{cg}-men (\ac{tss}). \ac{tw} were recruited during their treatment at the outpatient clinic of the department of psychiatry at the University of Münster. Before treatment and study inclusion all participants were carefully tested for chromosomal abnormalities such as Klinefelter syndrome, screened for personality disorders and other psychiatric comorbidities using the Structural Clinical Interview I and II according to DSM-IV criteria.

    Data of \ac{tw} and \ac{cg} were recorded under equal conditions (e.g. scanner, timeframe, study protocol, investigator), ruling out possible confounding of the classifier due to scanner variability. The \ac{tw} were in different treatment states, with 18 already treated with hormones (supplementary table\ref{tab:descript_stats_apllication}). Further details can be found in the original study ~\cite{Schoning2010}.

    \subsection{Image acquisition and structural preprocessing}
    Image acquisition and structural preprocessing followed previously published protocols for the \ac{mnc}\cite{Dannlowski2015b, Dannlowski2015a}, the FOR2107~\cite{Vogelbacher2018} and the BiDirect Cohort~\cite{Teuber2017}. A detailed description can be found in supplementary methods~\ref{subsec:supp_images_and_preprocessing}.

    \subsection{Analyses}

    \subsubsection{Multivariate analysis}
\begin{figure}[ht]
        \begin{center}
            \includegraphics[scale=0.5]{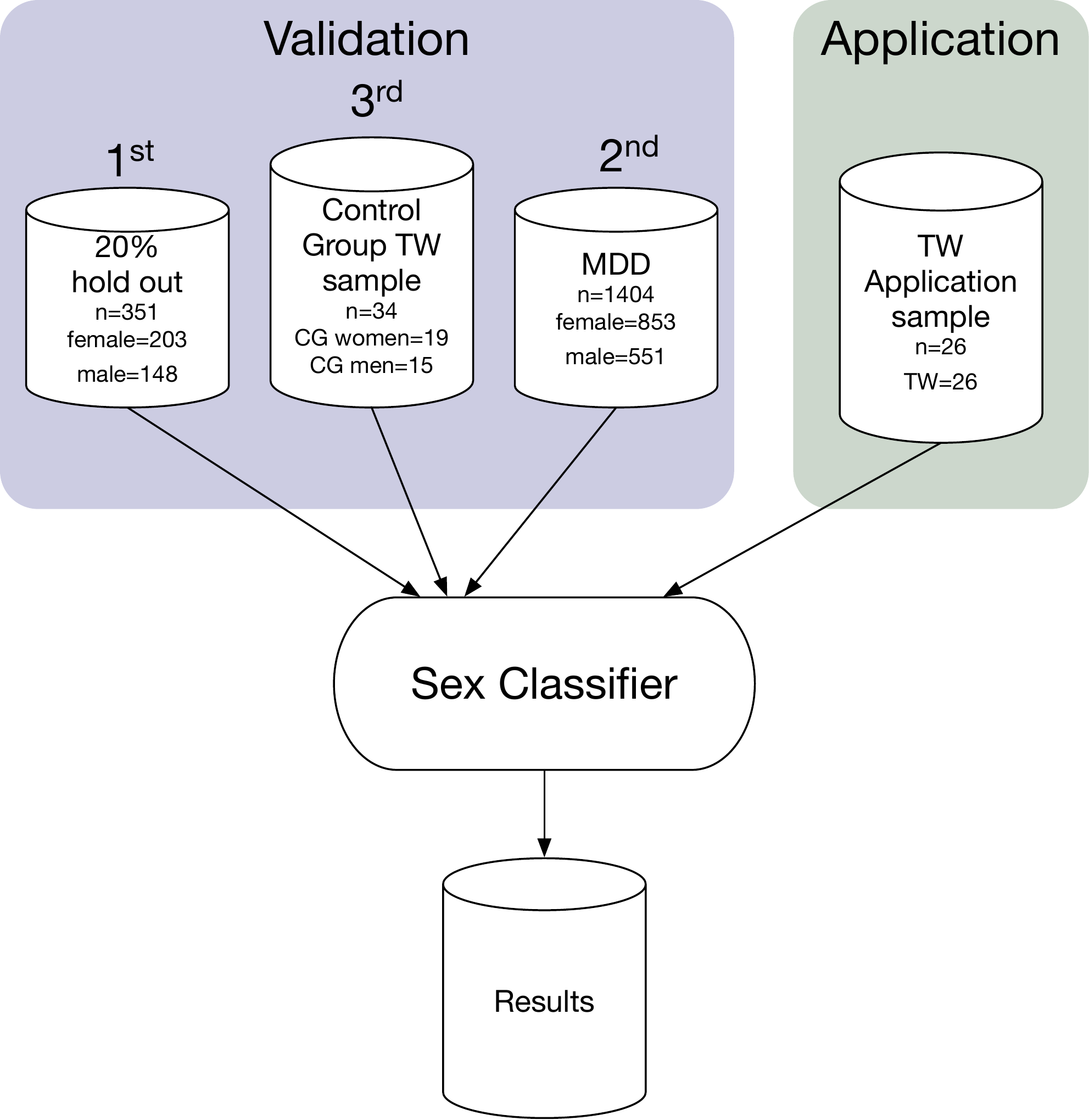}
        \end{center}
        \caption[Application of the trained classifier for biological sex prediction.]{Application of the trained classifier for biological sex prediction.

            \vspace{4pt}
            \setlength{\tabcolsep}{4pt}
            \begin{tabular}{llll}
                \hspace{-6pt} Abbreviations: & \acs{cg} &-& \acl{cg} \\
                & \acs{tw} &-& \acl{tw} \\
                & \acs{mdd} &-& \acl{mdd}
            \end{tabular}
        }
        \label{fig:application_trained_classifier}
    \end{figure}

    Individualized prediction of the biological sex was assessed with a \ac{svc}, implemented in the Scikit-learn toolbox~\cite{Pedregosa2012}. CAT12 whole-brain gray matter images were used as a classifier input~\cite{Gaser}. Gray matter images were resliced to a voxel size of $3 \times 3 \times 3$\,\si{\cubic\milli\meter}, to reduce dimensionality while preserving maximal localized morphometric differences.
    The training process was strictly separated from the evaluation, by selecting a random validation set of \SI{20}{\percent} ($N = 351$, $N_\text{female} = 219$, $N_\text{male} = 132$), which was not used during classifier training and testing. The remaining data set of $N = 1402$ subjects was balanced for sex with a random undersampling procedure ($N = 1218$, $N_\text{female} = 609$, $N_\text{male} = 609$), and used in a 10-fold split procedure resulting in balanced training sets of $N=1096$ subjects in each fold.
    A \ac{pca} was performed next, to further reduce the dimensionality of the data. The maximum number of principal components is limited to 1096, the number of subjects resulting from the 10-fold split. We carried out a Bayes-statistic-based hyperparameter optimization (Scikit-Optimize~\cite{Head2018}) for the SVC, nested in the 10-fold cross-validation. The parameter search included choice of the kernel (\ac{rbf} or linear), the $C$ parameter (\num{e-2} to \num{e2}, non-discrete log-scale), which influences penalties for misclassification, and the $\gamma$ parameter ($10^{-6}$ to 10, non-discrete log-sale), influencing the curvature of the decision boundary. In this iterative Bayes approach, a total of 100 parameter combinations were evaluated. Quality and classifier performance are reported by \ac{auc}. The classifier resulting from the best combination of hyperparameters was finally determined using our first validation set, the \SI{20}{\percent} drawn in advance from the original sample. To exclude potential effects of comorbid depression, this step was repeated with the sample of \ac{mdd} subjects, as a second validation sample (figure~\ref{fig:application_trained_classifier}).

    The final trained and validated classifier was then applied to the application sample with \acp{ti}. To test if classification results differ between \ac{cg}-men and \ac{tw} (same biological sex), we applied the \ac{tpr}. Since \ac{bacc} is a measure not applicable to one-group-only scenarios. Fisher’s exact test was used to clarify whether \ac{tpr} differs statistically between samples. Interpretation of \ac{tpr} is based on the hypothesis that \ac{tw} belong to the category of male biological sex.

    In order to achieve optimal generalization of our classifier, multiple scanners were deliberately incorporated. A specific correction for possible scanner effects was not intended. Instead, the purpose was to establish a classifier based on scanner invariant features given the large amount of training data and expected excellent classification performances. Comparison of the recognition rates between the individual scanners yielded no significant differences. Hence, an influence of the scanner on the classification results could not be detected - supporting our expectation (see supplementary table~\ref{tab:scanner_site_men_vs_women}). However, it should be pointed out that our data reveal a practically identical classification performance of the classifier trained on the multi-scanner training set (\SI{94.01}{\percent} \ac{bacc} in the first validation) to its application on the third validation-sample (\ac{cg} control group of the \ac{tw}-sample), using a different single scanner environment (\SI{94.03}{\percent} \ac{bacc}), suggesting that the classifier learned scanner independent features driving the classification performance.

    \subsubsection{Univariate analysis}
    The methodological details of the univariate analysis can be found in supplemental methods~\ref{subsec:analysis_univariat}.

    \section{Results}

    \subsection{Multivariate analysis}

    \subsubsection{\acl*{Cg}\acused{cg} training and first validation sample}
    The training of the classifier led to two results. The first result was the estimation of a hyperparameter set, determined with the Bayes optimization method. The hyperparameter optimization estimated an \ac{rbf} kernel, $C = 27.3$ and $\gamma = 2.4 \times 10^{-5}$ for the \ac{svm} as optimal approximation for the present problem.

\begin{figure}[ht]
        \begin{center}
            \input{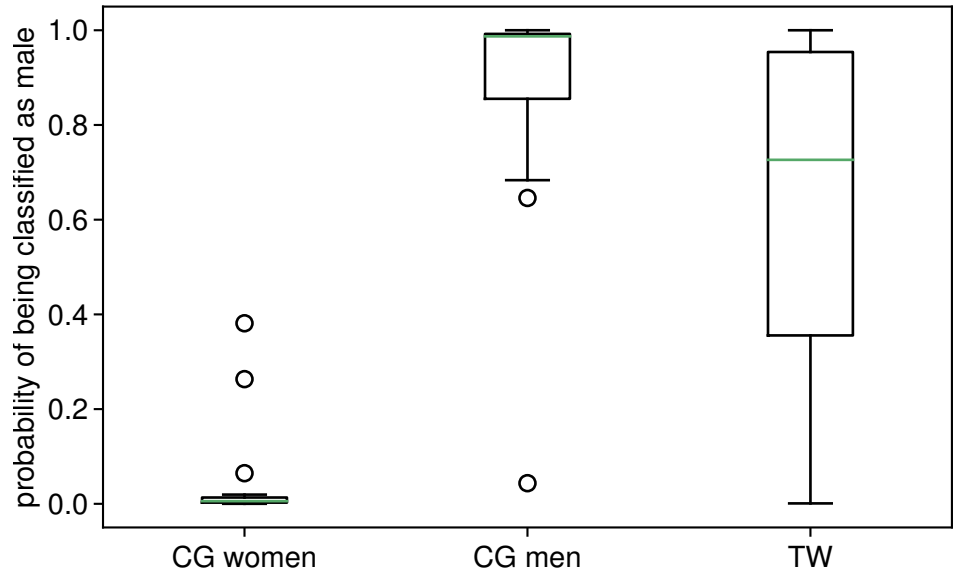}
        \end{center}
        \caption[]{Boxplot for the predicted probabilities of male sex based on the application sample and the third validation sample, including transgender and \acl{cg} individuals.

            \vspace{4pt}
            \setlength{\tabcolsep}{4pt}
            \begin{tabular}{llll}
                \hspace{-6pt} Abbreviations: & \acs{cg} &-& \acl{cg} \\
                & \acs{tw} &-& \acl{tw} \\
            \end{tabular}
        }
        \label{fig:boxplot_probibilities_for_male_classification}
    \end{figure}

    Based on the estimated hyperparameters, the second result was the classification outcome of the \SI{20}{\percent} validation set, which provided a performance indication for the trained classifier. The \acl{bacc} for the validation set classification was \SI{94.01}{\percent}.
    The confusion matrix (supplementary table~\ref{tab:results_first_application_set}) revealed that our classifier assigns the female biological sex (\ac{tpr} = \SI{99.9}{\percent}) more accurately than the male biological sex (\ac{tpr} = \SI{88.5}{\percent}). These results are visualized by a ROC curve, based on the probabilities for a classification as male (supplements figure~\ref{fig:boxplot_probibilities_for_male_classification}a), with a calculated \ac{auc} of \num{0.99}.

    \subsubsection{MDD second validation sample}
    To rule out that \ac{mdd} comorbidity had any influence on the classifier, we used a second validation set consisting of 1404 \ac{mdd} subjects (853 \ac{cg}-women, 551 \ac{cg}-men). Our classifier reached a \ac{bacc} of \SI{92.06}{\percent}, and a \ac{tpr} of \SI{86.93}{\percent} for \ac{cg}-men in this sample (supplementary table~\ref{tab:results_second_application_set}). The results of the classifier, the corresponding ROC curve (supplementary figure~\ref{fig:boxplot_probibilities_for_male_classification}d), and the \ac{auc} of 0.99 are similar to the results of the first validation set. Fisher’s exact test revealed no significant differences between the distribution of results of the first and second validation sample (supplementary table~\ref{tab:comparsion_fishers_test}).

    \subsubsection{Transgender application sample and \acl{cg} third validation sample}
    The \ac{bacc} for the third validation sample was \SI{94.03}{\percent} (\ac{cg} part of the \ac{tw}-sample). The \ac{tpr} for \ac{cg}-men was \SI{93.3}{\percent} and for \ac{cg}-women \SI{100.0}{\percent} (table~\ref{tab:classification_res_appl}). However, the \ac{tpr} for the \ac{tw} was remarkably low at \SI{56}{\percent} (supplementary table~\ref{tab:results_first_application_set}); see visualization by ROC curves (supplementary figure~\ref{fig:boxplot_probibilities_for_male_classification}b, c). The corresponding \ac{auc} differed as a function of group between 0.99 (\ac{cg}-men) and 0.95 (\ac{tw}). This difference in \ac{tpr} was significant, as Fisher's exact test showed a statistically significant difference between \ac{tpr} of \ac{cg}-men and \ac{tw} with hormone treatment (table~\ref{tab:classification_res_appl}). The output probabilities of the classifier are represented descriptively in figure~\ref{fig:boxplot_probibilities_for_male_classification}, as a box plot.

    \begin{table}[ht]

        \begin{center}
            \begin{tabular}{l|S[table-format=2.0]S[table-format=2.0]S[table-format=3.2]l}
                \multicolumn{1}{c}{}& \multicolumn{2}{c}{\textbf{N}} & \multicolumn{1}{c}{\textbf{\ac{tpr}}}&  \textbf{Fisher’s Exact Test} \\ \cmidrule(lr){2-3}
                \textbf{Group} & \multicolumn{1}{c}{\textbf{correct}} & \multicolumn{1}{c}{\textbf{total}} & \multicolumn{1}{c}{\textbf{in \%}} & \textbf{against \ac{cg}-men} \\
                \hline
                \acs{cg}-women & 19 & 19 & 100.00 & $p=1.0$ \\
                \acs{cg}-men& 14 & 15 & 93.33 & - \\
                \acs{tw} & 16 & 26 & 61.54 & $p=.028$ *** \\
                \acs{tw} (treatment naive) & 7 & 8 & 87.50 & $p=.999$ \\
                \acs{tw} (post \acs{cht}) & 9 & 18 & 50.00 & $p=.008$ *** \\

            \end{tabular}

        \end{center}
        \caption[Classification results in the application sample]{Classification results in the application sample. Classification results in percentage of true positive rate identified biological sex. \\

            \vspace{-6pt}
            \setlength{\tabcolsep}{4pt}
            \begin{tabular}{llll}
                \hspace{-6pt} Abbreviations: & \ac{tpr} &-& \acl{tpr} (sensitivity) \\
                & \acs{cg} &-& \acl{cg} \\
                & \acs{tw} &-& \acl{tw} \\
                & \acs{cht} &-& \acl{cht} \\
                & *** &-& indicates significance of the Fisher's Exact Test ($p < .05$)
            \end{tabular}
        }
        \label{tab:classification_res_appl}
    \end{table}

    \subsection{Univariate analysis}
\begin{figure}[ht]
        \begin{center}
            \includegraphics[scale=0.13]{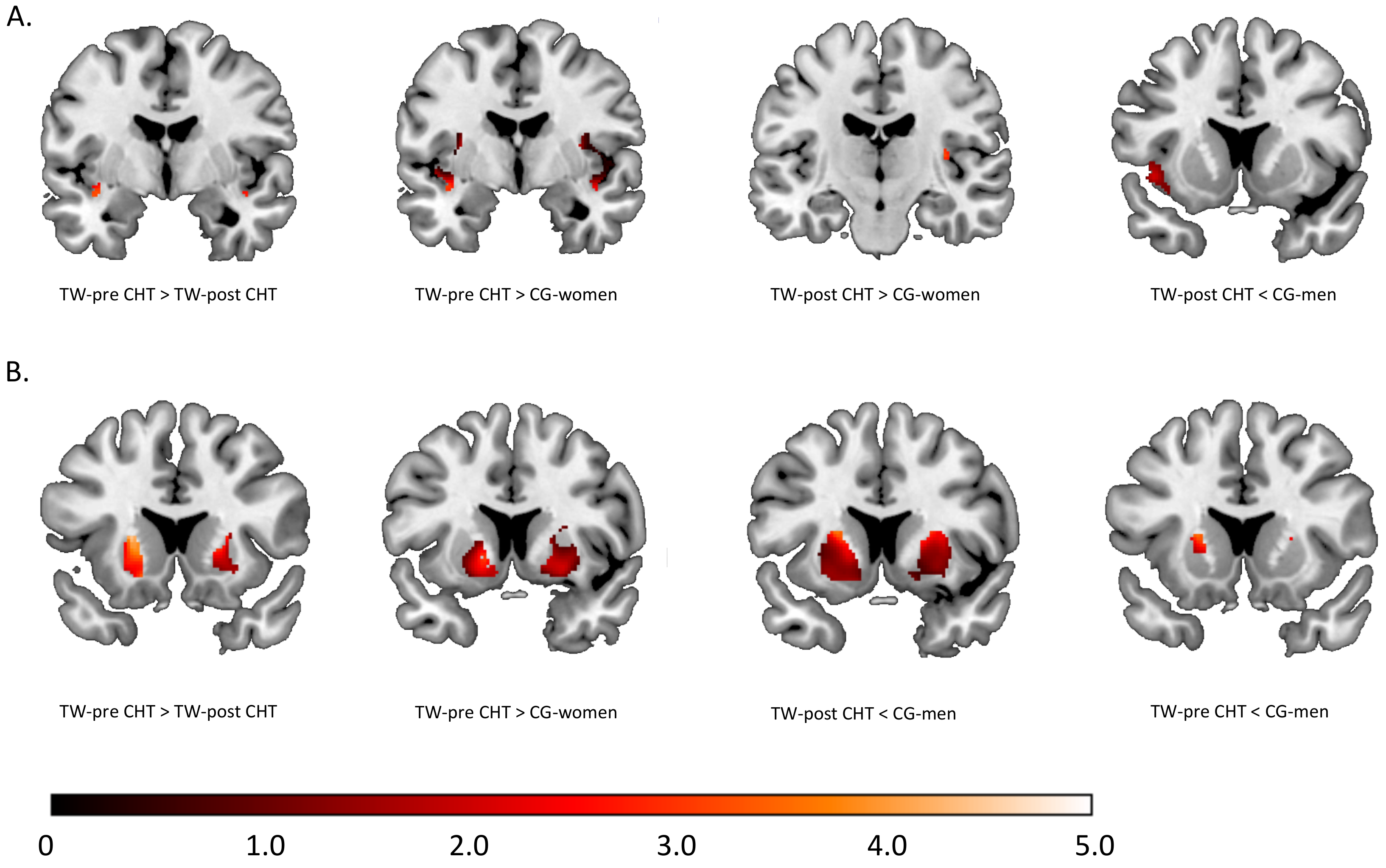}
        \end{center}
        \caption[]{Significant results of the univariate gray matter analysis. Color-bar represents t-values of the extracted clusters. Image shows the cluster at the respective peak voxel as reported in table~\ref{tab:results_univariat}. \\
            \textbf{A.} Alterations of the insula between groups (\acl{cg} men, \acl{cg} women and \acl{tw} before vs. after hormone treatment)\\
            \textbf{B.} Alterations of the putamen between groups (\acl{cg} men, \acl{cg} women \acl{tw} before vs. after hormone treatment)
        }
        \label{fig:results_univariate_gray_matter_analysis}
    \end{figure}

    The \acl{roi} analysis is summarized in table~\ref{tab:results_univariat} and figure~\ref{fig:results_univariate_gray_matter_analysis} (see coordinates and detailed statistics there). Briefly, using rigorous alpha correction, our analysis revealed no differences between \ac{tw}-post \ac{cht} and \ac{cg}-women in the bilateral putamen. In the insula, \ac{tw}-post \ac{cht} showed higher volume than \ac{cg}-women. \ac{tw}-post \ac{cht} and \ac{cg}-women both showed lower volume of the insula and putamen compared to \ac{cg}-men.
    In contrast, \ac{tw}-pre \ac{cht} showed larger volume in both \ac{roi} analyses compared to \ac{cg}-women. Interestingly, \ac{tw} pre \ac{cht} also showed higher volume in the putamen compared to \ac{cg}-men.
    \ac{tw} post-\ac{cht} showed lower volume of both regions of interest compared to \ac{tw} pre-\ac{cht} in both regions of interest. \ac{cg}-men showed larger volume in both regions of interest compared to \ac{cg}-women.
    Detailed results of our exploratory whole-brain analysis can be found in the supplementary table~\ref{tab:results_whole_brain_analysis}. Omitting \ac{tw} individuals with psychiatric comorbidities did not alter findings in general (see supplementary table~\ref{tab:roi_analysis_restricted_tw_sample} and~\ref{tab:results_whole_brain_analysis}). However, conclusions should be made with caution due to limited sample size.

    \begin{table}[]
        \begin{center}
            \begin{tabular}{l|llS[table-format=3.2]S[table-format=<0.3,add-integer-zero=false]S[table-format=4.0]S[table-format=+2.0]S[table-format=+2.0]S[table-format=+2.0]}

                \multicolumn{1}{c}{}                            & \multicolumn{1}{c}{\textbf{region}}             & & & & & \multicolumn{3}{c}{\textbf{MNI-space}}\\ \cmidrule(lr){7-9}
                \textbf{compared groups}                        & \textbf{of interest}        & \textbf{side} & \textbf{TFCE} & \textbf{p-FWE}  & \multicolumn{1}{c}{\textbf{k}}           & \multicolumn{1}{c}{x}          & \multicolumn{1}{c}{y}          & \multicolumn{1}{c}{z} \\ \hline
                \multirow{4}{*}{\ac{tw}-pre \textgreater{} \ac{tw}-post}  & \multirow{2}{*}{insula}     & L & 91.50 & .012 & 76 & -38 & -3 & -12        \\
                & & R & 54.96 & .033 & 23 & 32 & 10 & -16        \\ \cline{2-9}
                & \multirow{2}{*}{putamen}    & L & 466.55 & \textless{}.001 & 2005 & -21 & 16 & 8          \\
                & & R & 395.31 & \textless{}.001 & 1409 & 27 & -8 & 15         \\ \hline
                \multirow{4}{*}{\ac{tw}-pre \textgreater{} \ac{cg}-women} & \multirow{2}{*}{insula}     & L & 63.21 & \textless{}.001 & 1926 & -39 & -3 & -12        \\
                & & R & 52.58 & \textless{}.001 & 2299 & 34 & 15 & -10        \\ \cline{2-9}
                & \multirow{2}{*}{putamen}    & L & 274.31 & \textless{}.001 & 2381 & -21 & 10 & 23          \\
                & & R & 257.58 & \textless{}.001 & 2316 & 26 & -4 & 14         \\ \hline
                \multirow{2}{*}{\ac{tw}-pre \textgreater{} \ac{cg}-men}   & \multirow{2}{*}{putamen}    & L & 203.55 & \textless{}.001 & 892 & -21 & 15 & 9        \\
                & & R & 183.13 & \textless{}.001 & 576 & 28 & -3 & 15         \\ \hline
                \multirow{5}{*}{\ac{tw}-post \textless{} \ac{cg}-men}     & \multirow{3}{*}{insula}     & L & 38.96 & .005 & 303 & -42 & 14 & -6        \\
                & & L & 30.99 & .010 & 124 & -42 & -8 & 4        \\
                & & R & 21.37 & .001 & 131 & 30 & -18 & 20        \\ \cline{2-9}
                & \multirow{2}{*}{putamen}    & L & 100.64 & .001 & 1050 & -14 & 9 & -2          \\
                & & R & 70.60 & .001 & 1429 & 26 & 4 & -8         \\ \hline
                \ac{tw}-post \textless{} \ac{cg}-women & insula & R & 114.58 & .021 & 99 & 34 & -15 & 9        \\ \hline
                \multirow{5}{*}{\ac{cg}-men \textgreater{} \ac{cg}-women} & \multirow{3}{*}{insula}     & L & 49.7 & \textless{}.001 & 1199 & -44 & 14 & -8        \\
                & & L & 13.07 & .004 & 48 & -44 & -14 & 8        \\
                & & R & 109.23 & \textless{}.001 & 1789 & 39 & 16 & 3        \\ \cline{2-9}
                & \multirow{2}{*}{putamen}    & L & 81.13 & \textless{}.001 & 1972 & 26 & 6 & -4          \\
                & & R & 100.11 & \textless{}.001 & 1429 & 26 & 4 & -8         \\

            \end{tabular}
        \end{center}
        \caption[]{Results of the univariate gray matter region of interest analysis of the insula and putamen.\\
            Note. Table reports respective statistics of significant clusters of the group comparisons between transgender and \acl{cg} individuals. Clusters resulted from group comparisons corrected for total intracranial volume, age and sexual orientation.
            For reasons of brevity no results below a threshold of $k = 22$\,voxel have been reported.
            \\
            \\
            Abbreviations:

            \begin{tabular}{lll}
                \ac{tw} &-& \acl{tw} \\
                \acs{cg} &-& \acl{cg} \\
                pre/post &-& before/after hormone treatment \\
                L/R &-& left/right \\
                k &-& cluster size \\
                TFCE &-& Threshold-Free-Cluster-Enhancement with subsequent Family-Wise-Error-Correction.
            \end{tabular}
        }
        \label{tab:results_univariat}
    \end{table}

    \section{Discussion}
    In the present study, we developed an \ac{svm} using hyperparameter optimization resulting in an accurate classification of biological sex based on structural MRI images. The classifier, trained on a large training set of healthy \ac{cg} individuals, performed equally well in three independent validation samples of healthy \ac{cg} individuals, and \ac{cg} participants suffering from MDD. When applying the same classifier to structural MRI data of \ac{tw}, the \ac{svm} shows a much lower \ac{tpr}, resulting in significantly more misclassifications of the biological sex of \ac{tw} (male) in favor of their perceived gender (female). Moreover, the descriptive statistics of classification probabilities regarding \ac{tw} (figure~\ref{fig:boxplot_probibilities_for_male_classification}) indicate a pattern of prediction uncertainty that is not observable in \ac{cg}.

    Hence, our results shed light on two important aspects in biological psychiatry of \acp{ti}: 1) The impact of hormonal treatment on brain structure, 2) the separation of psychological distress (i.e. depression), hormonal treatment and trait characteristics of being a \ac{ti}.

    Our results replicate the finding that biological sex is increasingly misclassified in \acp{ti}, as previously described~\cite{Hoekzema2015, Baldinger-Melich2019}. This might encourage further investigations into the cause for increased misclassifications in \ac{tw}. Most notably and in contrast to previous studies, we could rule out that our findings are biased by comorbid depression and antidepressant medication. Given that the results of the first validation-sample of healthy \ac{cg}-participants were replicated in a large clinical sample of \ac{cg}-patients suffering from major depression, the classifier is reliable and robust to noise even from psychiatric disorders such as MDD and medication, which have been associated with structural brain changes~\cite{Redlich2018,Zaremba2018}.

    Our biological sex classifier shows a higher external validity than other biological sex classifiers. First, it has been tested on controls and MDD-patients, with high and very similar accuracy. Second, the \ac{svm} has been trained on large samples that have been collected at different sites. Hence, our \ac{svm} can be regarded as more generalizable while preserving performance and accuracy, indicating its robustness to noise.

    In the present work, we focused on the first application of this \ac{svm} on \ac{tw}. We observed that our \ac{svm} was increasingly inaccurate in \ac{tw}, compared to healthy \ac{cg} controls. The explorative analysis revealed that this inaccuracy was particularly increased in \ac{tw} who had hormonal treatment.

    Although our \ac{tw}-pre-\ac{cht} sample size was low, we aimed to differentiate structural brain alterations between \ac{tw}-pre and \ac{tw}-post-\ac{cht} as well as in comparison to \ac{cg}-women and -men. Our results show brain structural alterations dependent on the treatment state of \ac{tw}.
    Volumes of the insula and putamen were larger in \ac{tw}-pre-\ac{cht} than in \ac{cg}-women, while \ac{tw}-post-\ac{cht} showed lower volumes of the right insula compared to \ac{cg}-women.

    In comparison to \ac{cg}-men, \ac{tw}-pre-\ac{cht} showed larger volumes of the putamen, while \ac{tw}-post-\ac{cht} showed lower volumes of both insula and putamen. Thus, \ac{tw} independent of treatment state show brain structural alterations in our regions of interest in comparison to both, \ac{cg}-men and –women.

    Detailed analysis of \ac{tw}-pre compared to -post-\ac{cht} revealed a less pronounced pattern of structural brain alterations in \ac{tw}-post-\ac{cht} compared to \ac{cg}-women. Comparing \ac{tw}-pre with \ac{tw}-post-\ac{cht} revealed lower volume of \ac{tw}-post-\ac{cht} in both regions of interest, as well as the whole-brain analysis. This implies that \ac{cht} induces a further feminization of brain structure in \ac{tw}. This result fits with previous longitudinal studies that have shown reductions of cortical thickness in \ac{tw}-pre to post-\ac{cht} [26]. Structural and functional alterations of the insula have consistently been associated with \acp{ti} compared to \ac{cg}-individuals~\cite{Spizzirri2018, Savic2011, Kranz2015, Burke2018, Manzouri2019}. The insula is associated with body and self-perception. Behaviorally, \ac{tw} perceive an incoherence between their biological sex and perceived gender that is accompanied by altered insula activity in response to bodily sensations~\cite{Case2017}.

    Brain structural alterations of the putamen have been associated with \ac{tw} across multiple studies and independent of treatment state~\cite{Savic2011,Luders2009,Mueller2017}. We examined the putamen volume across different treatment states. Our study reveals that \ac{tw}-pre show a higher volume of the putamen compared to \ac{cg}-men and \ac{cg}-women, while \ac{tw}-post show lower volume of the putamen compared to \ac{cg}-men, but not to \ac{cg}-women. However, it remains unknown how \ac{cht} influences these structural alterations of \ac{tw}. Longitudinal examinations are required to reveal region specific structural alterations to estimate the impact of \ac{cht} of brain structure.

    Our combined univariate and multivariate approach revealed associations of \ac{cht} with lower accuracy in detecting the biological sex of \ac{tw}. Our results show that the brain structure of \ac{tw} aligns with neither their biological sex (male) nor their perceived gender (female). This implies that there is a biological basis for being transgender and thus, destigmatizes \acp{ti}. Further, this evidence can be used in psychoeducation during treatment of gender dysphoria. The diagnosis of gender dysphoria is new to DSM-5 to allow for treatment if \acp{ti} suffers from distress due to incoherence between perceived gender and biological sex. Our results could relieve distress in transgender patients in case of the experience of guilt or shame due to the discrepancy between biological sex and perceived gender.

    In line with this idea, hormonal processes, brain-structural development and the development of gender identity are intertwined~\cite{Nguyen2018a}. Intrauterine hormones drive the development of gender identity, rather than social learning processes~\cite{Bao2005, Swaab2009}. The male physical appearance is formed in the first trimester, due to effects of testosterone, and the female body develops due to the lack of androgens in this period~\cite{Shamim2000}. While the maturation of reproductive organs is more or less limited to the first trimester, brain development is continuing throughout pregnancy~\cite{Bao2011,VanGoozenS.H.Slabbekoorn2002}. Hormonal influences after the first trimester do not change the biological sex, but the experience of gender and thus might be responsible for the incoherence between biological and experienced sex. Since hormonal influences change gender perception as well as brain structure, \ac{cht} may lead to misclassifications in the \ac{tw}-group after treatment. Our univariate data indeed show that \ac{cht} is associated with structural brain alterations comparing \ac{tw}-pre and post-\ac{cht} to \ac{cg}-individuals. A previous study showed increased misclassification of biological sex even in untreated \ac{tw}~\cite{Hoekzema2015}, which we could not statistically support due to the small sample size of our untreated group ($N = 8$). Therefore, further studies should follow up on this effect, with higher sample sizes of untreated \ac{tw} to increase power. An extension of the design with a second control group (women with hormonal treatment) should be used to clarify whether misclassification is an effect of treatment only, due to the combination of being transgender and \ac{cht}.

    The present \ac{svc} provides a new tool for research in biological psychiatry. Prevalence of many psychiatric disorders is often higher for one biological sex than for the other. For example, prevalence in autism is higher for biological men than for biological women. Hence, it was hypothesized that female patients with autism might be similar in their brain structure to men. A previous study that developed a biological sex classifier using structural MRI scans and applied it to patients with autism~\cite{Ecker2017} indeed showed increased misclassifications of biological sex in female patients with autism. Therefore, biological sex misclassifications might point to involvement of aberrant biological sex development in the onset of such neurodevelopmental disorders. Future studies could use our trained classifier (\url{https://photon-ai.com/model_repo/bsc_mri}) to test for misclassifications in other clinical diagnoses with high gender imbalance in prevalence rates, such as eating disorders, substance use disorders, or anxiety disorders.

    \subsection{Limitations}
    Next to our training and validation strategy (visualized in figure~\ref{fig:flow_chart_classification}), a variety of other strategies exist such as repeated nested k-fold cross validation (see also~\cite{Baldinger-Melich2019}). The latter is an adequate means of choice in the absence of external validation-samples and produces robust estimates. However, even by preserving similar classification performances, we cannot rule out that other validation strategies could result in learning other patterns and therefore influence the prediction on \ac{tw} individuals. In addition, due to our small sample size of \ac{tw}, replication of the prediction failure of our \ac{svm} in \acp{ti} pre and post-\ac{cht} is needed. To verify that our effect is due to hormonal treatment, larger samples and studies in transgender men (biological sex female) are needed. Future studies should further dissect effects of gender dysphoria from depression, and effects of hormonal treatment from the state of being a TI\@.
    Finally, on the basis of the present data, we cannot draw firm conclusions on why the sensitivity of our classifier is greater towards the female. Further research is needed that investigates how classification performance in \ac{cg}-men and -women is associated with sex hormones.

    \section{Conclusion}
    In this study, we present a highly accurate biological sex classifier in \ac{cg}-individuals that shows a significantly decreased accuracy in \acp{ti} after \ac{cht}. Our results underline that the brain structure of \acp{ti} is similar to both, the brain structure of their perceived gender and biological sex. This implies that brain structure of \ac{tw} differs from both cg men and women. Based on our brain structural data, we suggest a dimensional rather than binary gender construct which will contribute to the destigmatization of \acp{ti}.

    \section*{Funding and Disclosure}
    This work was funded by the German Research Foundation (DFG, grant FOR2107 DA1151/5-1 and DA1151/5-2 to UD; SFB-TRR58, Projects C09 and Z02 to UD) and the Interdisciplinary Center for Clinical Research (IZKF) of the medical faculty of Münster (grant Dan3/012/17 to UD). The BiDirect Study is supported by a grant of the German Ministry of Research and Education (BMBF) to the University of Münster (01ER0816 and 01ER1506).

    Biomedical financial interests or potential conflicts of interest: Tilo Kircher received unrestricted educational grants from Servier, Janssen, Recordati, Aristo, Otsuka, neuraxpharm.

    The other authors (Claas Flint, Katharina Förster, Sophie A. Koser, Carsten Konrad, Pienie Zwitserlood, Klaus Berger, Marco Hermesdorf, Igor Nenadic, Axel Krug, Bernhard T. Baune, Katharina Dohm, Ronny Redlich, Nils Opel, Tim Hahn, Xiaoyi Jiang, Udo Dannlowski, Dominik Grotegerd) declare no conflicts of interest.

    \section*{Acknowledgements}
    \subsection*{FOR2107}
    This work is part of the German multicenter consortium “Neurobiology of Affective Disorders. A translational perspective on brain structure and function”, funded by the German Research Foundation (Deutsche Forschungsgemeinschaft DFG; Forschungsgruppe/Research Unit FOR2107).

    Principal investigators (PIs) with respective areas of responsibility in the FOR2107 consortium are:
    Work Package WP1, FOR2107/MACS cohort and brain imaging: Tilo Kircher (speaker FOR2107; DFG grant numbers KI 588/14-1, KI 588/14-2), Udo Dannlowski (co-speaker FOR2107; DA 1151/5-1, DA 1151/5-2), Axel Krug (KR 3822/5-1, KR 3822/7-2), Igor Nenadic (NE 2254/1-2), Carsten Konrad (KO 4291/3-1). WP2, animal phenotyping: Markus Wöhr (WO 1732/4-1, WO 1732/4-2), Rainer Schwarting (SCHW 559/14-1, SCHW 559/14-2). WP3, miRNA: Gerhard Schratt (SCHR 1136/3-1, 1136/3-2). WP4, immunology, mitochondriae: Judith Alferink (AL 1145/5-2), Carsten Culmsee (CU 43/9-1, CU 43/9-2), Holger Garn (GA 545/5-1, GA 545/7-2). WP5, genetics: Marcella Rietschel (RI 908/11-1, RI 908/11-2), Markus Nöthen (NO 246/10-1, NO 246/10-2), Stephanie Witt (WI 3439/3-1, WI 3439/3-2). WP6, multi method data analytics: Andreas Jansen (JA 1890/7-1, JA 1890/7-2), Tim Hahn (HA 7070/2-2), Bertram Müller-Myhsok (MU1315/8-2), Astrid Dempfle (DE 1614/3-1, DE 1614/3-2). CP1, biobank: Petra Pfefferle (PF 784/1-1, PF 784/1-2), Harald Renz (RE 737/20-1, 737/20-2). CP2, administration. Tilo Kircher (KI 588/15-1, KI 588/17-1), Udo Dannlowski (DA 1151/6-1), Carsten Konrad (KO 4291/4-1).

    Data access and responsibility: All PIs take responsibility for the integrity of the respective study data and their components. All authors and coauthors had full access to all study data.

    Acknowledgements and members by Work Package (WP):
    WP1: Henrike Bröhl, Katharina Brosch, Bruno Dietsche, Rozbeh Elahi, Jennifer Engelen, Sabine Fischer, Jessica Heinen, Svenja Klingel, Felicitas Meier, Tina Meller, Torsten Sauder, Simon Schmitt, Frederike Stein, Annette Tittmar, Dilara Yüksel (Dept. of Psychiatry, Marburg University). Mechthild Wallnig, Rita Werner (Core-Facility Brainimaging, Marburg University). Carmen Schade-Brittinger, Maik Hahmann (Coordinating Centre for Clinical Trials, Marburg). Michael Putzke (Psychiatric Hospital, Friedberg). Rolf Speier, Lutz Lenhard (Psychiatric Hospital, Haina). Birgit Köhnlein (Psychiatric Practice, Marburg). Peter Wulf, Jürgen Kleebach, Achim Becker (Psychiatric Hospital Hephata, Schwalmstadt-Treysa). Ruth Bär (Care facility Bischoff, Neunkirchen). Matthias Müller, Michael Franz, Siegfried Scharmann, Anja Haag, Kristina Spenner, Ulrich Ohlenschläger (Psychiatric Hospital Vitos, Marburg). Matthias Müller, Michael Franz, Bernd Kundermann (Psychiatric Hospital Vitos, Gießen). Christian Bürger, Fanni Dzvonyar, Verena Enneking, Stella Fingas, Janik Goltermann, Hannah Lemke, Susanne Meinert, Jonathan Repple, Kordula Vorspohl, Bettina Walden, Dario Zaremba (Dept. of Psychiatry, University of Münster). Harald Kugel, Jochen Bauer, Walter Heindel, Birgit Vahrenkamp (Dept. of Clinical Radiology, University of Münster). Gereon Heuft, Gudrun Schneider (Dept. of Psychosomatics and Psychotherapy, University of Münster). Thomas Reker (LWL-Hospital Münster). Gisela Bartling (IPP Münster). Ulrike Buhlmann (Dept. of Clinical Psychology, University of Münster).
    WP2: Marco Bartz, Miriam Becker, Christine Blöcher, Annuska Berz, Moria Braun, Ingmar Conell, Debora dalla Vecchia, Darius Dietrich, Ezgi Esen, Sophia Estel, Jens Hensen, Ruhkshona Kayumova, Theresa Kisko, Rebekka Obermeier, Anika Pützer, Nivethini Sangarapillai, Özge Sungur, Clara Raithel, Tobias Redecker, Vanessa Sandermann, Finnja Schramm, Linda Tempel, Natalie Vermehren, Jakob Vörckel, Stephan Weingarten, Maria Willadsen, Cüneyt Yildiz (Faculty of Psychology, Marburg University).
    WP4: Jana Freff, Silke Jörgens, Kathrin Schwarte (Dept. of Psychiatry, University of Münster). Susanne Michels, Goutham Ganjam, Katharina Elsässer (Faculty of Pharmacy, Marburg University). Felix Ruben Picard, Nicole Löwer, Thomas Ruppersberg (Institute of Laboratory Medicine and Pathobiochemistry, Marburg University).
    WP5: Helene Dukal, Christine Hohmeyer, Lennard Stütz, Viola Schwerdt, Fabian Streit, Josef Frank, Lea Sirignano (Dept. of Genetic Epidemiology, Central Institute of Mental Health, Medical Faculty Mannheim, Heidelberg University).
    WP6: Anastasia Benedyk, Miriam Bopp, Roman Keßler, Maximilian Lückel, Verena Schuster, Christoph Vogelbacher (Dept. of Psychiatry, Marburg University). Jens Sommer, Olaf Steinsträter (Core-Facility Brainimaging, Marburg University). Thomas W.D. Möbius (Institute of Medical Informatics and Statistics, Kiel University).
    CP1: Julian Glandorf, Fabian Kormann, Arif Alkan, Fatana Wedi, Lea Henning, Alena Renker, Karina Schneider, Elisabeth Folwarczny, Dana Stenzel, Kai Wenk, Felix Picard, Alexandra Fischer, Sandra Blumenau, Beate Kleb, Doris Finholdt, Elisabeth Kinder, Tamara Wüst, Elvira Przypadlo, Corinna Brehm (Comprehensive Biomaterial Bank Marburg, Marburg University).

    The FOR2107 cohort project (WP1) was approved by the Ethics Committees of the Medical Faculties, University of Marburg (AZ: 07/14) and University of Münster (AZ: 2014--422-b-S).

    \bibliographystyle{unsrt}
    \bibliography{references}

\begin{thebibliography}{10}

\bibitem{Zhou1995}
Jiang~Ning Zhou, Michel~A. Hofman, Louis~J.G. Gooren, and Dick~F. Swaab.
\newblock {A sex difference in the human brain and its relation to
  transsexuality}.
\newblock {\em Nature}, 378(6552):68--70, 1995.

\bibitem{Meyer-Bahlburg1996}
Heino~F.L. Meyer-Bahlburg, Rhoda~S. Gruen, Maria~I. New, Jennifer~J. Bell,
  Akira Morishima, Mona Shimshi, Yvette Bueno, Ileana Vargas, and Susan~W.
  Baker.
\newblock {Gender change from female to male in classical congenital adrenal
  hyperplasia}.
\newblock {\em Hormones and Behavior}, 30(4):319--332, dec 1996.

\bibitem{Mathews2009}
Greta~A. Mathews, Briony~A. Fane, Gerard~S. Conway, Charles~G.D. Brook, and
  Melissa Hines.
\newblock {Personality and congenital adrenal hyperplasia: Possible effects of
  prenatal androgen exposure}.
\newblock {\em Hormones and Behavior}, 55(2):285--291, feb 2009.

\bibitem{Bao2011}
Ai-Min Bao and Dick~F. Swaab.
\newblock {Sexual differentiation of the human brain: Relation to gender
  identity, sexual orientation and neuropsychiatric disorders}.
\newblock {\em Frontiers in Neuroendocrinology}, 32(2):214--226, apr 2011.

\bibitem{VanGoozen1995}
S~H {Van Goozen}, P~T Cohen-Kettenis, L~J Gooren, N~H Frijda, and N~E {Van de
  Poll}.
\newblock {Gender differences in behaviour: activating effects of cross-sex
  hormones.}
\newblock {\em Psychoneuroendocrinology}, 20(4):343--63, jan 1995.

\bibitem{Ruigrok2014}
Amber~N.V. Ruigrok, Gholamreza Salimi-Khorshidi, Meng-Chuan Lai, Simon
  Baron-Cohen, Michael~V. Lombardo, Roger~J. Tait, and John Suckling.
\newblock {A meta-analysis of sex differences in human brain structure}.
\newblock {\em Neuroscience {\&} Biobehavioral Reviews}, 39:34--50, feb 2014.

\bibitem{Cahill2006}
Larry Cahill.
\newblock {Why sex matters for neuroscience}.
\newblock {\em Nature Reviews Neuroscience}, 7(6):477--484, jun 2006.

\bibitem{McCarthy2011}
Margaret~M McCarthy and Arthur~P Arnold.
\newblock {Reframing sexual differentiation of the brain.}
\newblock {\em Nature neuroscience}, 14(6):677--83, jun 2011.

\bibitem{Spizzirri2018}
Giancarlo Spizzirri, F{\'a}bio Luis~Souza Duran, Tiffany~Moukbel
  Chaim-Avancini, Mauricio~Henriques Serpa, Mikael Cavallet, Carla Maria~Abreu
  Pereira, Pedro~Paim Santos, Paula Squarzoni, Naomi~Antunes da~Costa,
  Geraldo~F. Busatto, and Carmita Helena~Najjar Abdo.
\newblock {Grey and white matter volumes either in treatment-na{\"i}ve or
  hormone-treated transgender women: a voxel-based morphometry study}.
\newblock {\em Scientific Reports}, 8(1):736, dec 2018.

\bibitem{Joel2015}
Daphna Joel, Zohar Berman, Ido Tavor, Nadav Wexler, Olga Gaber, Yaniv Stein,
  Nisan Shefi, Jared Pool, Sebastian Urchs, Daniel~S Margulies, Franziskus
  Liem, J{\"u}rgen H{\"a}nggi, Lutz J{\"a}ncke, and Yaniv Assaf.
\newblock {Sex beyond the genitalia: The human brain mosaic.}
\newblock {\em Proceedings of the National Academy of Sciences of the United
  States of America}, 112(50):15468--73, dec 2015.

\bibitem{Luders2009}
Eileen Luders, Francisco~J S{\'a}nchez, Christian Gaser, Arthur~W Toga,
  Katherine~L Narr, Liberty~S Hamilton, and Eric Vilain.
\newblock {Regional gray matter variation in male-to-female transsexualism.}
\newblock {\em NeuroImage}, 46(4):904--7, jul 2009.

\bibitem{Savic2011}
I.~Savic and S.~Arver.
\newblock {Sex dimorphism of the brain in male-to-female transsexuals}.
\newblock {\em Cerebral Cortex}, 21(11):2525--2533, nov 2011.

\bibitem{Mueller2017}
Sven~C. Mueller, Lionel Landr{\'e}, Katrien Wierckx, and Guy T'Sjoen.
\newblock {A Structural Magnetic Resonance Imaging Study in Transgender Persons
  on Cross-Sex Hormone Therapy}.
\newblock {\em Neuroendocrinology}, 105(2):123--130, 2017.

\bibitem{Altinay2019}
Murat Altinay and Amit Anand.
\newblock {Neuroimaging gender dysphoria: a novel psychobiological model}.
\newblock {\em Brain Imaging and Behavior}, 2019.

\bibitem{Seiger2016}
Rene Seiger, Andreas Hahn, Allan Hummer, Georg~S. Kranz, Sebastian Ganger,
  Michael Woletz, Christoph Kraus, Ronald Sladky, Alexander Kautzky, Siegfried
  Kasper, Christian Windischberger, and Rupert Lanzenberger.
\newblock {Subcortical gray matter changes in transgender subjects after
  long-term cross-sex hormone administration}.
\newblock {\em Psychoneuroendocrinology}, 74:371--379, 2016.

\bibitem{Nguyen2018}
Hillary~B. Nguyen, A.~M. Chavez, E.~Lipner, Liisa Hantsoo, Sara~L. Kornfield,
  R.~D. Davies, and C.~N. Epperson.
\newblock {Gender-affirming hormone use in transgender individuals: Impact on
  behavioral health and cognition}.
\newblock {\em Current Psychiatry Reports}, 20(12), 2018.

\bibitem{Nguyen2018a}
Hillary~B. Nguyen, James Loughead, Emily Lipner, Liisa Hantsoo, Sara~L.
  Kornfield, and C.~Neill Epperson.
\newblock {What has sex got to do with it? the role of hormones in the
  transgender brain}.
\newblock {\em Neuropsychopharmacology}, 0:1--16, 2018.

\bibitem{WhiteHughto2016}
Jaclyn~M {White Hughto} and Sari~L Reisner.
\newblock {A systematic review of the effects of hormone therapy on
  psychological functioning and quality of life in transgender individuals}.
\newblock {\em Transgender Health}, 1(1):21--31, 2016.

\bibitem{Kranz2017}
Georg~S. Kranz, Rene Seiger, Ulrike Kaufmann, Allan Hummer, Andreas Hahn,
  Sebastian Ganger, Martin Tik, Christian Windischberger, Siegfried Kasper, and
  Rupert Lanzenberger.
\newblock {Effects of sex hormone treatment on white matter microstructure in
  individuals with gender dysphoria}.
\newblock {\em NeuroImage}, 150:60--67, apr 2017.

\bibitem{Mueller2017a}
Sven~C Mueller and Griet {De Cuypere}.
\newblock {Mechanisms of Psychiatric Illness Transgender Research in the 21st
  Century: A Selective Critical Review From a Neurocognitive Perspective}.
\newblock {\em Am J Psychiatry}, 174:1155--1162, 2017.

\bibitem{Hoekzema2015}
Elseline Hoekzema, Sebastian~E.E. Schagen, Baudewijntje~P.C. Kreukels, Dick~J.
  Veltman, Peggy~T. Cohen-Kettenis, Henriette {Delemarre-van de Waal}, and
  Julie Bakker.
\newblock {Regional volumes and spatial volumetric distribution of gray matter
  in the gender dysphoric brain}.
\newblock {\em Psychoneuroendocrinology}, 55:59--71, may 2015.

\bibitem{Baldinger-Melich2019}
Pia Baldinger-Melich, Maria F~Urquijo Castro, Dominic~B Seiger,
  Ren{\'e}and~Dwyer, Georg~S Kranz, Manfred Kl{\"o}bl, Ulrike Kaufmann,
  Christian Windischberger, Peter Falkai, Rupert Lanzenberger, and Nikolaos
  Koutsouleris.
\newblock {Sex Matters : A Multivariate Pattern Analysis of Sex- and
  Gender-Related Neuroanatomical Differences in Cis- and Transgender
  Individuals Using Structural Magnetic Resonance Imaging}.
\newblock {\em Cerebral Cortex}, pages 1--12, 2019.

\bibitem{Varoquaux2018}
Ga{\"e}l Varoquaux.
\newblock {Cross-validation failure: Small sample sizes lead to large error
  bars}.
\newblock {\em NeuroImage}, 180(Pt A):68--77, oct 2018.

\bibitem{Kranz2015}
Georg~S. Kranz, Wolfgang Wadsak, Ulrike Kaufmann, Markus Savli, Pia Baldinger,
  Gregor Gryglewski, Daniela Haeusler, Marie Spies, Markus Mitterhauser,
  Siegfried Kasper, and Rupert Lanzenberger.
\newblock {High-dose testosterone treatment increases serotonin transporer
  binding in transgender people}.
\newblock {\em Biological Psychiatry}, 78(8):525--533, oct 2015.

\bibitem{Burke2018}
Sarah~M Burke, Amir~H Manzouri, Cecilia Dhejne, Karin Bergstr{\"o}m, Stefan
  Arver, Jamie~D Feusner, and Ivanka Savic-Berglund.
\newblock {Testosterone effects on the brain in transgender men}.
\newblock {\em Cerebral Cortex}, 28(5):1582--1596, may 2018.

\bibitem{Zubiaurre-Elorza2014}
Leire Zubiaurre-Elorza, Carme Junque, Esther G{\'o}mez-Gil, and Antonio
  Guillamon.
\newblock {Effects of Cross-Sex Hormone Treatment on Cortical Thickness in
  Transsexual Individuals}.
\newblock {\em The Journal of Sexual Medicine}, 11(5):1248--1261, may 2014.

\bibitem{Simon2013}
Lajos Simon, Lajos~R. Koz{\'a}k, Vikt{\'o}ria Simon, P{\'a}l Czobor, Zsolt
  Unoka, {\'A}d{\'a}m Szab{\'o}, and G{\'a}bor Csukly.
\newblock {Regional grey matter structure differences between transsexuals and
  healthy controls—a voxel based morphometry study}.
\newblock {\em PLoS ONE}, 8(12):1--10, dec 2013.

\bibitem{Luders2012}
Eileen Luders, Francisco~J S{\'a}nchez, Duygu Tosun, David~W Shattuck,
  Christian Gaser, Eric Vilain, and Arthur~W Toga.
\newblock {Increased cortical thickness in male-to-female transsexualism}.
\newblock {\em Journal of behavioral and brain science}, 2(3):357--362, aug
  2012.

\bibitem{Wittchen1997}
H.-U. Wittchen, U.~Wunderlich, S.~Gruschwitz, and M.~Zaudig.
\newblock {\em {SKID-I. Strukturiertes Klinisches Interview f{\"u}r DSM-IV.}}
\newblock Hogrefe, G{\"o}ttingen, 1997.

\bibitem{Dannlowski2015}
U~Dannlowski, H~J Grabe, K~Wittfeld, J~Klaus, C~Konrad, D~Grotegerd, R~Redlich,
  T~Suslow, N~Opel, P~Ohrmann, J~Bauer, P~Zwanzger, I~Laeger, C~Hohoff,
  V~Arolt, W~Heindel, M~Deppe, K~Domschke, K~Hegenscheid, H~V{\"o}lzke,
  D~Stacey, H~{Meyer zu Schwabedissen}, H~Kugel, and B~T Baune.
\newblock {Multimodal imaging of a tescalcin (TESC)-regulating polymorphism
  (rs7294919)-specific effects on hippocampal gray matter structure}.
\newblock {\em Molecular Psychiatry}, 20(3):398--404, mar 2015.

\bibitem{Teuber2017}
Anja Teuber, Benedikt Sundermann, Harald Kugel, Wolfram Schwindt, Walter
  Heindel, Jens Minnerup, Udo Dannlowski, Klaus Berger, and Heike Wersching.
\newblock {MR imaging of the brain in large cohort studies: feasibility report
  of the population- and patient-based BiDirect study}.
\newblock {\em European Radiology}, 27(1):231--238, jan 2017.

\bibitem{Kircher2018}
Tilo Kircher, Markus W{\"o}hr, Igor Nenadic, Rainer Schwarting, Gerhard
  Schratt, Judith Alferink, Carsten Culmsee, Holger Garn, Tim Hahn, Bertram
  M{\"u}ller-Myhsok, Astrid Dempfle, Maik Hahmann, Andreas Jansen, Petra
  Pfefferle, Harald Renz, Marcella Rietschel, Stephanie~H. Witt, Markus
  N{\"o}then, Axel Krug, and Udo Dannlowski.
\newblock {Neurobiology of the major psychoses: a translational perspective on
  brain structure and function—the FOR2107 consortium}.
\newblock {\em European Archives of Psychiatry and Clinical Neuroscience},
  pages 1--14, sep 2018.

\bibitem{Vogelbacher2018}
Christoph Vogelbacher, Thomas~W.D. M{\"o}bius, Jens Sommer, Verena Schuster,
  Udo Dannlowski, Tilo Kircher, Astrid Dempfle, Andreas Jansen, and Miriam~H.A.
  Bopp.
\newblock {The Marburg-M{\"u}nster Affective Disorders Cohort Study (MACS): A
  quality assurance protocol for MR neuroimaging data}.
\newblock {\em NeuroImage}, 172:450--460, may 2018.

\bibitem{Teismann2014}
Henning Teismann, Heike Wersching, Maren Nagel, Volker Arolt, Walter Heindel,
  Bernhard~T Baune, J{\"u}rgen Wellmann, Hans-Werner Hense, and Klaus Berger.
\newblock {Establishing the bidirectional relationship between depression and
  subclinical arteriosclerosis – rationale, design, and characteristics of
  the BiDirect Study}.
\newblock {\em BMC Psychiatry}, 14(1):174, dec 2014.

\bibitem{Schoning2010}
Sonja Sch{\"o}ning, Almut Engelien, Christine Bauer, Harald Kugel, Anette
  Kersting, Cornelia Roestel, Pienie Zwitserlood, Martin Pyka, Udo Dannlowski,
  Wolfgang Lehmann, Walter Heindel, Volker Arolt, and Carsten Konrad.
\newblock {Neuroimaging differences in spatial cognition between men and
  male-to-female transsexuals before and during hormone therapy}.
\newblock {\em Journal of Sexual Medicine}, 7(5):1858--1867, 2010.

\bibitem{Dannlowski2015b}
Udo Dannlowski, Harald Kugel, Dominik Grotegerd, Ronny Redlich, Janina Suchy,
  Nils Opel, Thomas Suslow, Carsten Konrad, Patricia Ohrmann, Jochen Bauer,
  Tilo Kircher, Axel Krug, Andreas Jansen, Bernhard~T Baune, Walter Heindel,
  Katharina Domschke, Andreas~J Forstner, Markus~M N{\"o}then, Jens Treutlein,
  Volker Arolt, Christa Hohoff, Marcella Rietschel, and Stephanie~H Witt.
\newblock {NCAN cross-disorder risk variant is associated with limbic gray
  matter deficits in healthy subjects and major depression}.
\newblock {\em Neuropsychopharmacology}, 40(11):2510--2516, oct 2015.

\bibitem{Dannlowski2015a}
Udo Dannlowski, Harald Kugel, Dominik Grotegerd, Ronny Redlich, Nils Opel,
  Katharina Dohm, Dario Zaremba, Anne Gr{\"o}gler, Juliane Schwieren, Thomas
  Suslow, Patricia Ohrmann, Jochen Bauer, Axel Krug, Tilo Kircher, Andreas
  Jansen, Katharina Domschke, Christa Hohoff, Pienie Zwitserlood, Markus
  Heinrichs, Volker Arolt, Walter Heindel, and Bernhard~T. Baune.
\newblock {Disadvantage of social sensitivity: Interaction of oxytocin receptor
  genotype and child maltreatment on brain structure}.
\newblock {\em Biological Psychiatry}, pages 1--8, 2015.

\bibitem{Pedregosa2012}
Fabian Pedregosa, Ga{\"e}l Varoquaux, Alexandre Gramfort, Vincent Michel,
  Bertrand Thirion, Olivier Grisel, Mathieu Blondel, Peter Prettenhofer, Ron
  Weiss, Vincent Dubourg, Jake Vanderplas, Alexandre Passos, David Cournapeau,
  Matthieu Brucher, Matthieu Perrot, and {\'E}douard Duchesnay.
\newblock {Scikit-learn: Machine Learning in Python}.
\newblock {\em Journal of Machine Learning Research}, 12:2825--2830, 2012.

\bibitem{Gaser}
Christian Gaser.
\newblock {Manual Computational Anatomy Toolbox- cat12}.

\bibitem{Head2018}
Tim Head, MechCoder Mik, Gilles Louppe, Iaroslav Shcherbatyi, Fcharras,
  Z{\'e}and Hvass-Labs Vin{\'i}cius, Cmmalone, Christopher Schr{\"o}der,
  Nel215, Nuno Campos, Young Todd, Carlosdanielcsantos, Stefano Cereda,
  Schwabedal, Thomas Fan, Rene-rex, Kejia~(KJ) Shi, and Justus.
\newblock scikit-optimize, 2018.

\bibitem{Redlich2018}
Ronny Redlich, Nils Opel, Christian B{\"u}rger, Katharina Dohm, Dominik
  Grotegerd, Katharina F{\"o}rster, Dario Zaremba, Susanne Meinert, Jonathan
  Repple, Verena Enneking, Elisabeth Leehr, Joshua~Joscha B{\"o}hnlein, Lena
  Winters, Neele Frob{\"o}se, Sophia Thrun, Julia Emtmann, Walter Heindel,
  Harald Kugel, Volker Arolt, Georg Romer, Christian Postert, and Udo
  Dannlowski.
\newblock {The limbic system in youth depression: Brain structural and
  functional alterations in adolescent in-patients with severe depression}.
\newblock {\em Neuropsychopharmacology}, 43(3):546--554, feb 2018.

\bibitem{Zaremba2018}
Dario Zaremba, Katharina Dohm, Ronny Redlich, Dominik Grotegerd, Robert
  Strojny, Susanne Meinert, Christian B{\"u}rger, Verena Enneking, Katharina
  F{\"o}rster, Jonathan Repple, Nils Opel, Bernhard~T. Baune, Pienie
  Zwitserlood, Walter Heindel, Volker Arolt, Harald Kugel, and Udo Dannlowski.
\newblock {Association of brain cortical changes with relapse in patients with
  major depressive disorder}.
\newblock {\em JAMA Psychiatry}, 75(5):484--492, 2018.

\bibitem{Manzouri2019}
A~Manzouri and I~Savic.
\newblock {Possible Neurobiological Underpinnings of Homosexuality and Gender
  Dysphoria}.
\newblock {\em Cerebral Cortex}, 29(5):2084--2101, 2019.

\bibitem{Case2017}
Laura~K. Case, David Brang, Rosalynn Landazuri, Pavitra Viswanathan, and
  Vilayanur~S. Ramachandran.
\newblock {Altered white matter and sensory response to bodily sensation in
  female-to-male transgender individuals}.
\newblock {\em Archives of Sexual Behavior}, 46(5):1223--1237, jul 2017.

\bibitem{Bao2005}
Ai-Min Bao, Andon Hestiantoro, Eus J. W.~Van Someren, Dick~F. Swaab, and
  Jiang-Ning Zhou.
\newblock {Colocalization of corticotropin-releasing hormone and oestrogen
  receptor- in the paraventricular nucleus of the hypothalamus in mood
  disorders}.
\newblock {\em Brain}, 128(6):1301--1313, 2005.

\bibitem{Swaab2009}
Dick~F Swaab and Alicia Garcia-Falgueras.
\newblock {Sexual differentiation of the human brain in relation to gender
  identity and sexual orientation}.
\newblock {\em Functional neurology}, 24(1):17--28, 2009.

\bibitem{Shamim2000}
Waqar Shamim, Mohammed Yousufuddin, Ameet Bakhai, Andrew~J.S. Coats, and
  John~W. Honour.
\newblock {Gender differences in the urinary excretion rates of cortisol and
  androgen metabolites}.
\newblock {\em Annals of Clinical Biochemistry}, 37(6):770--774, 2000.

\bibitem{VanGoozenS.H.Slabbekoorn2002}
D.~{van Goozen, S. H., Slabbekoorn}, L.~J. Gooren, G.~Sanders, and P.~T.
  Cohen-Kettenis.
\newblock {Organizing and activating effects of sex hormones in homosexual
  transsexuals. Behavioral neurosci}.
\newblock {\em Behavioral Neuroscience}, 116(6):982--988, 2002.

\bibitem{Ecker2017}
Christine Ecker, Derek~S. Andrews, Christina~M. Gudbrandsen, Andre~F. Marquand,
  Cedric~E. Ginestet, Eileen~M. Daly, Clodagh~M. Murphy, Meng~Chuan Lai,
  Michael~V. Lombardo, Amber~N.V. Ruigrok, Edward~T. Bullmore, John Suckling,
  Steven~C.R. Williams, Simon Baron-Cohen, Michael~C. Craig, and Declan~G.M.
  Murphy.
\newblock {Association between the probability of autism spectrum disorder and
  normative sex-related phenotypic diversity in brain structure}.
\newblock {\em JAMA Psychiatry}, 74(4):329--338, 2017.

\bibitem{Tzourio-Mazoyer2002}
N~Tzourio-Mazoyer, B~Landeau, D~Papathanassiou, F~Crivello, O~Etard,
  N~Delcroix, B~Mazoyer, M~Joliot, and B~Mazoyer.
\newblock {Automated anatomical labeling of activations in SPM using a
  macroscopic anatomical parcellation of the MNI MRI single-subject brain}.
\newblock {\em NeuroImage}, 15(1):273--289, jan 2002.

\end{thebibliography}

    \newpage
    \appendix
    \pagenumbering{roman}

    \lhead{\textsc{Supplements}}

    \chead{Biological sex misclassification in \acl{tw}}
    \begin{center}
        \LARGE{\textsc{Supplemental Materials}} \\ \Large{Biological sex classification with structural MRI data shows increased misclassification in \acl{tw}}
    \end{center}

    \setcounter{equation}{1}
    \setcounter{figure}{0}
    \setcounter{table}{0}
    \setcounter{page}{1}

    \renewcommand\thefigure{\thesection.\arabic{figure}}
    \renewcommand\thetable{\thesection.\arabic{table}}

    \section{Supplemental Tables and Figures}
    \begin{table}[ht]
        \renewcommand{\arraystretch}{1.5}
        \begin{center}
            \begin{tabular}{lr|lccc}
                \multicolumn{3}{c}{} & {\textbf{female}} & {\textbf{male}} & {\textbf{significance test}} \\ \cline{2-6}
                \textbf{Münster Neuro-} & \acs{hc} & Quantity & 278 & 388 &  \\
                \textbf{imaging Cohort} & & Age & 36.4 (11.7) & 35.8 (12.6) & $F(1,665)=0.4917$, $p=.480$\\ \cline{3-6}
                & \acs{mdd} & Quantity &121 & 164 \\
                & & Age & 36.4 (11.7) & 35.8 (12.6) & $F(1,665)=0.4917$, $p=.480$\\
                & & BDI & 23.4 (9.5) & 26.5 (10.9) & $F(1,274)=6.212$, $p=.013$\\
                & & HDRS-17 & 18.5 (3.8) & 19.5 (4.5) & $F(1,187)=2.739$, $p=.100$\\
                & & IQ (MWBT) & 111.5 (13.3) & 110.9 (14.1) & $F(1,267)=2.739$, $p=.765$\\  \cline{2-6}
                \textbf{BiDirect Study} & \acs{hc} & Quantity & 217 & 217 \\
                & & Age & 51.3 (8.1) & 53.0 (8.0) & $F(1,433)=4.481$, $p=.035$ \\ \cline{3-6}
                & \acs{mdd} & Quantity & 235 & 356 \\
                & & Age & 48.1 (7.4) & 49.6 (7.3) & $F(1,590)=5.612$,  $p=.019$ \\
                & & HDRS-17 & 12.6 (6.8) & 14.3 (6.5) & $F(1,587)=10.160$, $p=.002$ \\ \cline{2-6}
                \textbf{FOR2107 Study} & \acs{hc} & Quantity & 246 & 407 \\
                & & Age & 32.6 (11.4) & 32.6 (13.0) & $F(1,652)=0.001$, $p=.976$ \\ \cline{3-6}
                & \acs{mdd} & Quantity & 195 & 333 \\
                & & Age & 36.6 (13.8) & 37.8 (13.4) & $F(1,527)=0.719$, $p=.397$  \\
                & & HDRS-17 & 8.9 (6.7) & 8.5 (6.8) & $F(1,525)=0.414$, $p=.520$  \\
            \end{tabular}
        \end{center}
        \vspace{15pt}
        \caption[]{Descriptive statistics of the trainings and validation samples. Table reports means and standard deviations of the individual cohorts used for the training of the \acl{svm}. Significance test was a univariate ANOVA without covariates. \\
            \\
            \setlength{\tabcolsep}{4pt}
            \begin{tabular}{llll}
                \hspace{-7pt} Abbreviations: & \acs{hc} &-& \acl{hc} \\
                & \acs{mdd} &-& \acl{mdd} \\
                & Age &-& age in years \\
            \end{tabular}
        }
        \label{tab:descrip_stats_train_validation}
    \end{table}

    \begin{table}[ht]
        \renewcommand{\arraystretch}{1.2}
        \begin{center}
            \begin{tabular}{lccccc}
                & \ac{cg} men & \ac{cg} women & \ac{tw} pre & \ac{tw} post &  \\
                & $(N=15)$ & $(N=19)$ & $(N=8)$ & $(N=18)$ & significance test \\ \hline
                Age & 34 & 32 & 33.9 & 33.1 & \multirow{2}{*}{$F(3,56)=0.190$, $p=.991$}\\
                & (8.6) & (6.3) & (14.1) & (31.3) \\ \hline
                Highest & 4.9 & 4.8 & 5.0 & 5.1 & \multirow{2}{*}{$F(3,56)=0.802$, $p=.498$}\\
                Education & (0.9) & (0.8) & (0.0) & (0.5)

            \end{tabular}
        \end{center}
        \vspace{15pt}
        \caption[]{Descriptive statistics of the application sample (transgender and \acl{cg} individuals). Table reports means and standard deviations of the transgender individuals and controls from a similar measurement period used for the test of the \acl{svm} in \ac{tw}. Significance test was univariate ANOVA without covariates. From 15 out of 29 \ac{tw} were requested whether they had a depressive episode and 8 from the 15 \ac{tw} indicated that they had a depressive episode.
            \emph{Highest Education} was measured according to educational attainment in numbers from 1 = special school to 6 = universal degree. \\
            \\
            \setlength{\tabcolsep}{4pt}
            \begin{tabular}{llll}
                \hspace{-7pt} Abbreviations: & \acs{tw} &-& \acl{tw} \\
                & pre/post &-& before/after \acl{cht} \\
                & Age &-& age in years \\
            \end{tabular}
        }
        \label{tab:descript_stats_apllication}
    \end{table}

    \begin{table}[ht]

        \begin{subtable}[c]{0.55\textwidth}
            \renewcommand{\arraystretch}{1.2}
            \begin{tabular}{rr|c|c}
                & \multicolumn{1}{c}{} & \multicolumn{2}{c}{actual} \\
                & group & female & male \\ \cline{2-4}
                \multirow{4}{*}{\rotatebox{90}{predicted}} & \multirow{2}{*}{female} & 202 & 17\\
                & & (\ac{tpr} = \SI{99.9}{\percent}) & (\ac{tnr} = \SI{11.5}{\percent})\\ \cline{2-4}
                & \multirow{2}{*}{male} & 1 & 131 \\
                & & (\ac{tnr} = \SI{0.1}{\percent}) & (\ac{tpr} = \SI{88.5}{\percent})\\
            \end{tabular}
        \end{subtable}
        \begin{subtable}[c]{0.2\textwidth}
            \begin{tabular}{rl}
                \multicolumn{2}{c}{} \\
                Accuracy &    \SI{94.87}{\percent} \\
                Balanced Accuracy &     \SI{94.01}{\percent} \\
                Precision &    \SI{99.24}{\percent} \\
                Recall & \SI{88.51}{\percent} \\
                F1-Score & 0.9357 \\

            \end{tabular}
        \end{subtable}
        \vspace{15pt}
        \caption[]{Results of the validation set ($N = 351$; $N_{\text{male}} = 148$; $N_{\text{female}} = 203$). Classification results in absolute numbers and percentage of accurately identified biological sex.

            \vspace{2pt}
            \setlength{\tabcolsep}{4pt}
            \begin{tabular}{llll}
                \hspace{-6pt} Abbreviations: & \acs{tpr} &-& \acl{tpr} (sensitivity) \\
                & \ac{tnr} &-& \acl{tnr} (specificity) \\
            \end{tabular}
        }
        \label{tab:results_validatons_set}
    \end{table}

    \begin{table}[ht]

        \begin{subtable}[c]{0.65\textwidth}
            \renewcommand{\arraystretch}{1.2}
            \begin{tabular}{rr|c|c|c}
                & \multicolumn{1}{c}{} & \multicolumn{3}{c}{actual} \\
                & group & \ac{cg} women & \ac{cg} men & \ac{tw}\\ \cline{2-5}
                \multirow{4}{*}{\rotatebox{90}{predicted}} & \multirow{2}{*}{female} & 19 & 1 & 10\\
                & & (\ac{tpr} = \SI{100.0}{\percent}) & (\ac{tnr} = \SI{6.7}{\percent}) & (\ac{tnr} = \SI{38.5}{\percent})\\ \cline{2-5}
                & \multirow{2}{*}{male} & 0 & 14 & 16\\
                & & (\ac{tnr} = \SI{0.0}{\percent}) & (\ac{tpr} = \SI{93.3}{\percent}) & (\ac{tpr} = \SI{61.5}{\percent})\\
            \end{tabular}
        \end{subtable}
        \begin{subtable}[c]{0.2\textwidth}
            \begin{tabular}{rl}
                \multicolumn{2}{l}{The following metrics are} \\
                \multicolumn{2}{l}{related to the \ac{cg} groups only:} \\
                Accuracy &    \SI{94.12}{\percent} \\
                Balanced Accuracy &\SI{96.67}{\percent} \\
                F1-Score & \num{0.9655} \\
            \end{tabular}
        \end{subtable}
        \vspace{15pt}
        \caption[]{Results of the application set ($N = 60$; $N_{\text{\ac{cg} men}} = 15$; $N_{\text{\ac{cg} women}} = 19$; $N_\text{\ac{tw}} = 26$). Classification results in absolute numbers and percentage of accurately identified biological sex.

            \vspace{2pt}
            \setlength{\tabcolsep}{4pt}
            \begin{tabular}{llll}
                \hspace{-6pt} Abbreviations: & \acs{tpr} &-& \acl{tpr} (sensitivity) \\
                & \acs{tnr} &-& \acl{tnr} (specificity) \\
                & \acs{cg} &-& \acl{cg} \\
                & \acs{tw} &-& \acl{tw} \\
            \end{tabular}
        }
        \label{tab:results_first_application_set}
    \end{table}

    \begin{table}[ht]

        \begin{subtable}[c]{0.55\textwidth}
            \renewcommand{\arraystretch}{1.2}
            \begin{tabular}{rr|c|c}
                & \multicolumn{1}{c}{} & \multicolumn{2}{c}{actual} \\
                & group & female & male \\ \cline{2-4}
                \multirow{4}{*}{\rotatebox{90}{predicted}} & \multirow{2}{*}{female} & 829 & 72\\
                & & (\ac{tpr} = \SI{97.2}{\percent}) & (\ac{tnr} = \SI{13.1}{\percent})\\ \cline{2-4}
                & \multirow{2}{*}{male} & 24 & 479 \\
                & & (\ac{tnr} = \SI{2.8}{\percent}) & (\ac{tpr} = \SI{86.9}{\percent})\\
            \end{tabular}
        \end{subtable}
        \begin{subtable}[c]{0.2\textwidth}
            \begin{tabular}{rl}
                \multicolumn{2}{c}{} \\
                Accuracy &    \SI{93.16}{\percent} \\
                Balanced Accuracy &     \SI{92.06}{\percent} \\
                Precision &    \SI{95.23}{\percent} \\
                Recall & \SI{86.93}{\percent} \\
                F1-Score & 0.9206 \\

            \end{tabular}
        \end{subtable}
        \vspace{15pt}
        \caption[]{Results of the second validation set  ($N = 1404$; $N_{\text{male}} = 551$; $N_{\text{female}} = 853$). Classification results in absolute numbers and percentage of accurately identified biological sex.

            \vspace{2pt}
            \setlength{\tabcolsep}{4pt}
            \begin{tabular}{llll}
                \hspace{-6pt} Abbreviations: & \acs{tpr} &-& \acl{tpr} (sensitivity) \\
                & \acs{tnr} &-& \acl{tnr} (specificity) \\
            \end{tabular}
        }
        \label{tab:results_second_application_set}
    \end{table}

    \begin{table}[ht]
        \renewcommand{\arraystretch}{1.2}
        \begin{center}
            \begin{tabular}{l|S[table-format=3.0]S[table-format=2.0]|S[table-format=3.0]S[table-format=2.0]|S[table-format=0.4]}
                \multicolumn{1}{c}{}& \multicolumn{2}{c}{validation set 1} & \multicolumn{2}{c}{validation set 2} & {Fisher’s Exact Test} \\
                & $N_\text{correct}$ & $N_\text{incorrect}$ & $N_\text{correct}$ & $N_\text{incorrect}$ & {$p$-value} \\ \hline
                \ac{cg} women & 131 & 17 & 479 & 72 & .7386 \\
                \ac{cg} men & 202 & 1 & 829 & 24 & .9955 \\

            \end{tabular}
        \end{center}
        \vspace{15pt}
        \caption[]{Classification results in the application sample. Comparison of the distribution of classification results between the first and second validation sets, using Fisher's exact test. (\ac{cg} - \acl{cg})
        }
        \label{tab:comparsion_fishers_test}
    \end{table}

    \begin{table}
        \centering

        \begin{tabular}{l|lrrrrrr}
            \multicolumn{1}{c}{}& & & & & \multicolumn{3}{c}{\textbf{MNI-Space}} \\ \cmidrule(lr){6-8}
            \textbf{compared groups} & \textbf{region of interest} & \textbf{TFCE} & \textbf{p-FWE} & \textbf{k} & \textbf{x} & \textbf{y} & \textbf{z} \\ \hline
            \multirow{9}{*}{\ac{tw}-pre > \ac{tw}-post} & R medial cingulate cortex & \multirow{2}{*}{1643.35} & \multirow{2}{*}{.005} & \multirow{2}{*}{15117} & \multirow{2}{*}{27} & \multirow{2}{*}{-9} & \multirow{2}{*}{16} \\
            & and caudate nucleus \\
            & L caudate nucleus & 1514.98 & .005 & 13228 & -22 & 18 & 12 \\
            & L precentral and middle frontal gyrus & 766.08 & .029 & 470 & -36 & 4 & 54 \\
            & L precuneus & 743.23 & .029 & 651 & -14 & -56 & 52 \\
            & L postcentral gyrus & 668.59 & .040 & 410 & -63 & -14 & 39 \\
            & R cerebellum & 632.84 & .045 & 505 & 48 & -54 & -27 \\
            & R cerebellum & 611.18 & .047 & 258 & 40 & -38 & -46 \\ \hline
            \multirow{4}{*}{\ac{tw}-pre > \ac{cg}-women} & L precuneus, R medial cingulate Cortex, & \multirow{2}{*}{982.71} & \multirow{2}{*}{<.001} & \multirow{2}{*}{108742} & \multirow{2}{*}{22} & \multirow{2}{*}{-3} & \multirow{2}{*}{15} \\
            & L + R lingual gyrus \\
            & L cerebellum & 328.77 & .003 & 4657 & -30 & -33 & -50 \\
            & R precentral, frontal inferior gyrus & 127.46 & .041 & 360 & 62 & 16 & 27 \\ \hline
            \multirow{2}{*}{\ac{tw}-post > \ac{cg}-women} & R calcarine/lingual gyrus, precuneus & 1312.08 & <.001 & 3875 & 4 & -50 & 3 \\
            & L cuneus, superior occipital gyrus & 605.11 & .023 & 498 & -10 & -96 & 36 \\
            \ac{tw}-post < \ac{cg}-women & R postcentral gyrus & 1057.71 & .026 & 562 & 45 & -21 & 32 \\ \hline
            \multirow{6}{*}{\ac{tw}-pre > \ac{cg}-men} & L caudate nucleus, putamen, hippocampus & 745.18 & .009 & 3018 & -21 & 18 & 9 \\
            & R caudate nucleus, putamen & 713.97 & .010 & 2336 & 28 & -4 & 16 \\
            & R Precuneus, Mid Cingulum & 673.63 & .013 & 3185 & 8 & -28 & 39 \\
            & L Pre-, Postcentral & 654.18 & .015 & 822 & -69 & -15 & -32 \\
            & R Hippocampus, Parahippocampus & 567.93 & .028 & 713 & 20 & -8 & -30 \\
            & R Calcarine, Lingual gyrus & 528.98 & .002 & 416 & 26 & -75 & 3 \\ \hline
            \multirow{3}{*}{\ac{tw}-post < \ac{cg}-men} & L middle temporal lobe, cerebellum & \multirow{2}{*}{939.61} & \multirow{2}{*}{<.001} & \multirow{2}{*}{111736} & \multirow{2}{*}{32} & \multirow{2}{*}{-56} & \multirow{2}{*}{-38} \\
            & R middle temporal lobe, cerebellum \\
            & R precentral and frontal inferior gyrus & 260.12 & .016 & 628 & 57 & 6 & 12 \\
        \end{tabular}
        \vspace{14pt}
        \caption[]{Results of the whole-brain analysis. For reasons of brevity only significant clusters $> k = 300$\,voxel are reported, we did not calculate a contrast comparing \acl{cg} men and women. The reported significant clusters resulted from group comparisons within a full factorial model corrected for total intracranial volume, age and sexual orientation. \\
            \\
            \setlength{\tabcolsep}{4pt}
            \begin{tabular}{llll}
                \hspace{-6pt} Abbreviations: & \acs{tw} &-& \acl{tw} \\
                & \acs{cg} &-& \acl{cg} \\
                & pre/post &-& before/after \acl{cht} \\
                & L/R &-& left/right \\
                & k &-& cluster size \\
                & TFCE &-& Threshold-Free-Cluster-Enhancement with subsequent Family-Wise-Error-Correction. \\
            \end{tabular}
        }
        \label{tab:results_whole_brain_analysis}
    \end{table}

    \begin{table}[ht]
        \begin{center}
            \begin{tabular}{l|S[table-format=1.0]S[table-format=1.0]}
                \multicolumn{1}{c}{}& {\ac{tw} pre} & {\ac{tw} post} \\
                & {($N=2$)} & {($N=8$)} \\ \hline
                Substance abuse & 1 & 1 \\
                Eating Disorder & {-} & 1 \\
                Obsessive-compulsive Disorder & {-} & 1 \\
                Anxiety Disorder & 1 & 1 \\
                Major Depressive Disorder & 2 & 6 \\
            \end{tabular}
        \end{center}
        \caption[Comorbidities of transgeder individuals.]{Comorbidities of transgender individuals. Psychiatric comorbidities were reported for 10 \acl{tw}. \\

            \vspace{-6pt}
            \setlength{\tabcolsep}{4pt}
            \begin{tabular}{llll}
                \hspace{-7pt} Abbreviations: & \ac{tw} &-& \acl{tw} \\
                & pre/post &-& before/after \acl{cht} \\
            \end{tabular}
        }
        \label{tab:psychiatric_comorbidities}
    \end{table}

    \begin{table}[ht]
        \renewcommand{\arraystretch}{1.2}
        \begin{center}
            \begin{tabular}{rl|S[table-format=2.0]S[table-format=1.0]|S[table-format=3.0]S[table-format=2.0]|S[table-format=3.0]S[table-format=1.0]|S[table-format=2.0]S[table-format=1.0]}

                \multicolumn{2}{c}{} & \multicolumn{4}{c}{\textbf{first validation sample}} & \multicolumn{4}{c}{\textbf{second validation sample}} \\
                \multicolumn{2}{c}{} & \multicolumn{2}{c}{{\ac{cg}-men}} & \multicolumn{2}{c|}{{\ac{cg}-women}} & \multicolumn{2}{c}{{\ac{cg}-men}} & \multicolumn{2}{c}{{\ac{cg}-women}}\\
                \multicolumn{2}{c}{} & {correct} & {incorrect} & {correct} & {incorrect} & {correct} & {incorrect} & {correct} & {incorrect}\\ \cline{2-10}
                \multirow{4}{*}{\rotatebox{90}{\textbf{scanner}}} & {FOR2107 Münster} & 14 & 2 & 26 & 0 & 66 & 3 & 109 & 4\\
                & {FOR2107 Marburg} & 25 & 6 & 54 & 1 & 107 & 19 & 213 & 7\\
                & {MNC Münster} & 44 & 7 & 84 & 0 & 103 & 18 & 160 & 4\\
                & {BiDirect Münster} & 48 & 2 & 38 & 0 & 203 & 32 & 347 & 9\\ \cline{2-10}
                & \textbf{Fisher's Exact Test} & \multicolumn{2}{c|}{$p=.131$} & \multicolumn{2}{c|}{$p=.586$}  & \multicolumn{2}{c|}{$p=.101$}&\multicolumn{2}{c}{$p=.882$}  \\
            \end{tabular}
        \end{center}
        \caption[Classification results separated by scanner features and site]{Classification results separated by scanner features and site. Table shows frequencies of correctly and incorrectly classified \ac{cg}-women and –men for all validation samples that used different scanners. The third validation sample was measured with the same MRI-sequence as MNC Münster. \\

            \vspace{-6pt}
            \setlength{\tabcolsep}{4pt}
            \begin{tabular}{llll}
                \hspace{-6pt} Abbreviations: & \acs{cg} &-& \acl{cg} \\
            \end{tabular}
        }
        \label{tab:scanner_site_men_vs_women}
    \end{table}

    \begin{table}[ht]
        \renewcommand{\arraystretch}{1.2}
        \begin{center}
            \begin{tabular}{rl|S[table-format=3.2]S[table-format=<0.3, add-integer-zero=false]S[table-format=4.0]S[table-format=+2.0]S[table-format=+2.0]S[table-format=+2.0]}
                \multicolumn{5}{c}{} & \multicolumn{3}{c}{\textbf{MNI-space}} \\ \cmidrule(lr){6-8}
                \multicolumn{2}{c|}{} & \textbf{TFCE} & \textbf{p-FWE} & {\textbf{k}} & {\textbf{x}} & {\textbf{y}} & {\textbf{z}} \\ \cline{2-8}
                \textbf{Insula} & \ac{tw}-pre > post & & {n.s.} \\
                & \ac{tw}-pre < post & & {n.s.} \\
                & \ac{tw}-pre > \ac{cg}-women & 13.70 & .002 & 643 & 40 & 20 & -8\\
                & & 2.14 & .014 & 107 & -24 & 21 & -15 \\
                & \ac{tw}-pre < \ac{cg}-women & 121.60 & .001 & 471 & -40 & -3 & 0\\
                & \ac{tw}-pre > \ac{cg}-men & 73.19 & \textless{}.001 & 2267 & -32 & -8 & 12 \\
                & & 19.90 & \textless{}.001 & 1271 & 38 & -20 & 3\\
                & & 3.18 & .001 & 620& 46 & 22 & -4\\
                & \ac{tw}-pre < \ac{cg}-men & & {n.s.}  \\
                & \ac{tw}-post > \ac{cg}-women & 4.76 & \textless{}.001 & 1058 & 40 & 21 & 8 \\
                & & 1.11 & .001 & 479 & -33 & 26 & 4\\
                & \ac{tw}-post < \ac{cg}-women & 357.70 & \textless{}.001 & 524 & -40 & -8 & 2\\
                & \ac{tw}-post > \ac{cg}-men & 24.46 & \textless{}.001 & 2427 & -32 & 3 & 12\\
                & & 9.54 & \textless{}.001 & 1309 & 48 & -10 & 3\\
                & & 5.41 & \textless{}.001 & 230 & 45 & 24 & -3\\
                & & 2.11 & \textless{}.001 & 200 & 33 & 16 & -21\\
                & \ac{tw}-post < \ac{cg}-men & & {n.s.}\\ \cline{2-8}
                \textbf{Putamen}  & \ac{tw}-pre > post & 445.23 & .005 & 1179 & -21 & 16 & 8 \\
                & \ac{tw}-pre < post & & {n.s.}\\
                & & 355.78 & .008 & 519 & 24 & 6 & 15 \\
                & \ac{tw}-pre > \ac{cg}-women & 99.88 & \textless{}.001 & 1848 & 26 & 6 & 15 \\
                & & 84.36 & \textless{}.001 & 2006 & -20 & 21 & -3  \\
                & \ac{tw}-pre > \ac{cg}-men & 266.62 & \textless{}.001 & 2351 & -21 & 15 & 8\\
                & & 155.27 & \textless{}.001 & 2130 & 28 & -3 & 15 \\
                & \ac{tw}-pre < \ac{cg}-men & & {n.s.}\\
                & \ac{tw}-post > \ac{cg}-women & & {n.s.}\\
                & \ac{tw}-post < \ac{cg}-women & 134.84 & .002 & 324 & -15 & 9 & 3 \\
                & \ac{tw}-post > \ac{cg}-men & 24.46 & \textless{}.001 & 272 & -32 & 3 & 12 \\
                & & 2.10 & \textless{}.001 & 215 & -28 & 6 & 10 \\
                & & 0.52 & \textless{}.001 & 119 & 34 & 2 & 4\\
                & \ac{tw}-post < \ac{cg}-men & & {n.s.}\\

            \end{tabular}
        \end{center}
        \caption[ROI analysis in the restricted transgender sample without psychiatric comorbidities]{\ac{roi} analysis in the restricted transgender sample without psychiatric comorbidities. Table shows significant clusters $> k = 100$~voxel for reasons of brevity. Transgender individuals that showed comorbidities according to the structured clinical interview based on DSM-IV-criteria were excluded from the analysis leaving $N = 6$ \ac{tw}-pre and $N = 10$ \ac{tw}-post. \\

            \vspace{-6pt}
            \setlength{\tabcolsep}{4pt}
            \begin{tabular}{llll}
                \hspace{-6pt} Abbreviations: & \ac{tw} &-& \acl{tw} \\
                & \acs{cg} &-& \acl{cg} \\
                & pre-/post &-& before/after \acl{cht} \\
                & TFCE &-& statistic of the non-parametric approach using threshold-free cluster  \\
                & & & enhancement \\
                & k &-& clustersize \\
                & n.s. &-& not significant, $k > 100$ \\
            \end{tabular}
        }
        \label{tab:roi_analysis_restricted_tw_sample}
    \end{table}
    \begin{figure}[ht]
        \begin{center}
            \includegraphics[scale=0.5]{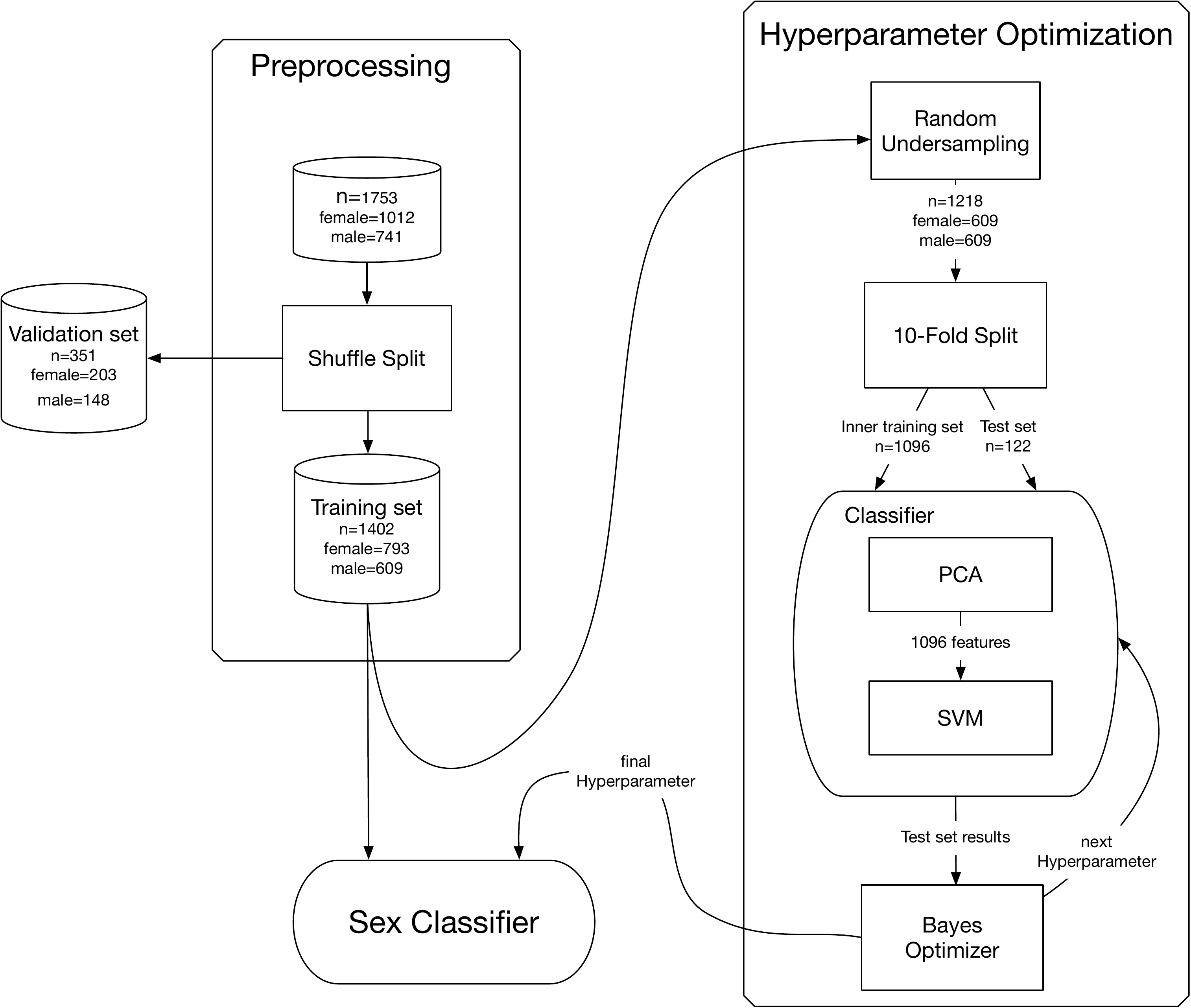}
        \end{center}
        \caption[]{Overview of training, validation and application procedure of the biological sex classifier.

            \vspace{4pt}
            \setlength{\tabcolsep}{4pt}
            \begin{tabular}{llll}
                \hspace{-6pt}
                Abbreviations: & \ac{svm} &-& \acl{svm} \\
                & \acs{pca} &-& \acl{pca} \\
            \end{tabular}
        }
        \label{fig:flow_chart_classification}
    \end{figure}

    \begin{figure}
        \captionsetup[subfigure]{justification=justified,singlelinecheck=false}
        \begin{subfigure}[b]{0.45\textwidth}
            \input{img/roc_curves_classification_as_male_1st_validation_sample.pgf}
            \caption[]{First validation sample: \ac{cg} men}
        \end{subfigure}
        \hspace{5.0mm}
        \begin{subfigure}[b]{0.45\textwidth}
            \input{img/roc_curves_classification_as_male_3rd_validation_sample.pgf}
            \caption[]{Third validation sample: \ac{cg} men}
        \end{subfigure}

        \vspace{5.0mm}

        \begin{subfigure}[b]{0.45\textwidth}
            \input{img/roc_curves_classification_as_male_application_sample.pgf}
            \caption[]{Application sample: \ac{tw} (biological sex male)}
        \end{subfigure}
        \hspace{5.0mm}
        \begin{subfigure}[b]{0.45\textwidth}
            \input{img/roc_curves_classification_as_male_2nd_validation_sample.pgf}
            \caption[]{Second Validation sample: MDD \ac{cg} men}
        \end{subfigure}
        \caption[]{Receiver Operation Characteristics for the classification as biological sex male for \ac{cg}-men in first, second and third validation sample as well \acl{tw} of our application sample. \\
            \\
            \vspace{-8pt}
            \setlength{\tabcolsep}{4pt}
            \begin{tabular}{llll}
                \hspace{-7pt} Abbreviations: & \acs{cg} &-& \acl{cg} \\
                & \acs{tw} &-& \acl{tw} \\
                & \acs{mdd} &-& \acl{mdd} \\
                & \acs{auc} &-& \acl{auc} \\
            \end{tabular}
        }
        \label{fig:roc_curves_classification_as_male}
    \end{figure}

    \FloatBarrier

    \section{Supplemental Methods}

    \subsection{Images and Structural Preprocessing}
    \label{subsec:supp_images_and_preprocessing}
    T1-weighted high-resolution anatomical images of the MNC and TSS were acquired at a 3T MRI (Gyroscan Intera 3T, Philips Medical Systems, the Netherlands) using a three-dimensional fast gradient echo sequence (turbo field echo), repetition time = \SI{7.4}{\milli\second}, echo time = \SI{3.4}{\milli\second}, flip angle = \ang{9}, two signal averages, inversion pre-pulse every \SI{814.5}{\milli\second}, acquired over a field of view of 256 (feet-head) $\times$ 204 (anterior-posterior) $\times$ \SI{160}{\cubic\milli\meter} (right-left), frequency encoding in feet to head direction, phase encoding in anterior-posterior and right-left direction, reconstructed to voxels of $0.5 \times 0.5 \times 0.5$\,\si{\cubic\milli\meter}~\cite{Dannlowski2015b, Dannlowski2015a}.

    The 3D T1-weighted turbo field echo images of the BD study were collected in the same scanner with repetition time = \SI{7.26}{\milli\second}, echo time = \SI{3.56}{\milli\second}, \ang{9} flip angle, 160 sagittal slices, matrix dimension $256 \times 256$, FOV = $256 \times 256$\,\si{\squared\milli\meter}, \SI{2}{\milli\meter} slice thickness (reconstructed to \SI{1}{\milli\meter}) resulting in a voxel size of $1 \times 1 \times 1$\,\si{\cubic\milli\meter}.
    The FOR2107 study was conducted at two different sites~\cite{Vogelbacher2018}. In Münster, data were collected with a 3T Siemens PRISMA using 3D T1-weighted magnetization prepared rapid acquisition gradient echo (MPRAGE) with repetition time = \SI{1900}{\milli\second}, echo time = \SI{2.28}{\milli\second}, inversion time = \SI{900}{\milli\second}, \ang{8} flip angle, 192 sagittal slices, \SI{0}{\milli\meter} slice gap, resulting in a voxel size of $1 \times 1 \times 1$\,\si{\cubic\milli\meter}. In Marburg, data were collected in a 3T Siemens Magnetom Trio Tim syngo MR B17 using a 3D T1-weighted magnetization prepared rapid acquisition gradient echo (MPRAGE) with repetition time = \SI{1900}{\milli\second}, echo time = \SI{2.26}{\milli\second}, inversion time = \SI{900}{\milli\second}, \ang{9} flip angle, 176 sagittal slices, \SI{0.5}{\milli\meter} slice lap, resulting in a voxel size of $1 \times 1 \times 1$\,\si{\cubic\milli\meter}.
    The structural images were preprocessed using the CAT12-toolbox~\cite{Gaser} (version r1184) in all four cohorts (MNC, FOR2107, BiDirect, TSS) following published protocols. Briefly, images were bias-corrected, tissue classified and normalized to MNI-space~\cite{Tzourio-Mazoyer2002}. For the univariate analysis, images were additionally smoothed with a Gaussian kernel of \SI{8}{\milli\meter} full width half maximum (FWHM). Absolute threshold masking with a threshold value of 0.1 was used for all univariate second-level analyses (\url{http://www.neuro.uni-jena.de/cat12/CAT12-Manual.pdf}). We carefully checked the sample for poor image quality detected by visual inspection and with the check homogeneity using covariance function implemented in CAT12.

    \subsection{Univariate analysis}
    \label{subsec:analysis_univariat}
    The \ac{tss} sample (\ac{tw} group and matched \ac{cg} controls, supplementary table~\ref{tab:descript_stats_apllication}) were used in the univariate analysis. Statistical parametric mapping (SPM12, Wellcome Trust Centre for Neuroimaging, London, \url{http://www.fil.ion.ucl.ac.uk/spm/}) was used for univariate gray matter analysis. The putamen and insula were defined as a priori \acp{roi} using the aal-atlas~\cite{Tzourio-Mazoyer2002} implemented in the Wake Forest University Pickatlas (\url{http://fmri.wfubmc.edu/software/PickAtlas}). We investigated the relationship between groups (\ac{cg}-men, -women, \ac{tw}-pre and -post \ac{cht}) and gray-matter volume with an ANCOVA, with age, total intracranial volume and sexual orientation as nuisance regressors in all analyses. Sexual orientation was indicated by the participants as a continuous variable (0-100, 0 indicating homosexuality, 50 indicating bisexuality and 100 indicating heterosexuality). Participants could choose each number between 0 and 100. The terminology was chosen according to the natal biological sex of \ac{tw}, i.e. homosexuality indicated sexual interest in men.
    We calculated a priori defined t-contrasts according to our hypothesis: \ac{cg}-men > women, \ac{cg}-men > \ac{tw}-pre, \ac{tw}-pre > \ac{cg}-women, \ac{cg}-men > \ac{tw}-post, \ac{tw}-post > \ac{cg}-women and \ac{tw} pre > \ac{tw} post. An additional whole brain analysis further explored possible regions with volume differences between the groups. Family-wise error correction with $p < .05$ was used in order to correct for alpha inflation. To determine statistical significance of putative clusters in each of the two bilateral \acp{roi} (insula, putamen) and the whole brain analysis, after applying the non-parametric approach of Threshold-Free Cluster Enhancement as implemented in the TFCE toolbox (\url{http://dbm.neuro.uni-jena.de/tfce}, version 167) we conducted 5000 permutations per test.

    \section{Supplemental Results \& Discussion}
    Basal testosterone was significantly associated with classification as male ($r = .437$, $p = .014$) across both \ac{tw} groups, while progesterone only showed a tendency towards a significant association with classification accuracy ($r = .274$, $p = .09$). Looking at \ac{tw}-pre-\ac{cht} and -post-\ac{cht} separately, \ac{tw}-pre \ac{cht} showed no significant association ($r = .218$, $p = .302$), while \ac{tw}-post-\ac{cht} showed a tendency towards a significant association between basal testosterone concentration and classification as male ($r = .392$, $p =.06$).
    The classification performance was significantly associated with the amount of basal testosterone measured in \ac{tw}. However, this association alone is not sufficient to draw any conclusions on the effect of \ac{cht} on classification performance, since we did not directly measure the amount of estradiol and anti-androgens that were taken. For instance, it is possible that \ac{tw}-pre-\ac{cht} contain lower levels of basal testosterone without taking any medication which might also increase the feeling of incongruence between natal sex and perceived gender (for a review~\cite{Nguyen2018a}).

    \section{Abbreviations}
    \begin{acronym}
        \acro{auc}[AUC]{area under the ROC curve}
        \acro{bacc}[BACC]{Balanced Accuracy}
        \acro{bd}[BD]{BiDirect study}
        \acro{cg}[CG]{cisgender}
        \acro{Cg}[CG]{Cisgender}
        \acro{cht}[CHT]{cross-sex hormone treatment}
        \acro{hc}[HC]{Healthy Control}
        \acro{mdd}[MDD]{major depressive disorder}
        \acro{mnc}[MNC]{Münster Neuroimaging Cohort}
        \acro{pca}[PCA]{principal-component analysis}
        \acro{rbf}[RBF]{radial basis function}
        \acro{roi}[ROI]{region of interest}
        \acroplural{roi}[ROIs]{regions of interest}
        \acro{svc}[SVC]{support-vector classifier}
        \acro{svm}[SVM]{support-vector machine}
        \acro{ti}[TI]{transgender individual}
        \acro{Ti}[TI]{Transgender individual}
        \acro{tnr}[TNR]{true negative rate}
        \acro{tpr}[TPR]{true positive rate}
        \acro{tss}[TSS]{Transgender study}
        \acro{tw}[TW]{transgender woman}
        \acroplural{tw}[TWs]{transgender women}
    \end{acronym}
\end{document}